\pdfoutput=1
\documentclass[11pt]{article}
\usepackage[final]{acl}
\usepackage{times}
\usepackage{latexsym}
\usepackage{algpseudocode}
\usepackage[T1]{fontenc}
\usepackage[utf8]{inputenc}
\usepackage{microtype}
\usepackage{inconsolata}
\usepackage{graphicx}
\usepackage{subcaption} 
\usepackage{multirow}
\usepackage{color} 
\usepackage{hyperref}
\usepackage{tcolorbox} 
\usepackage{enumitem}

\definecolor{brown}{RGB}{139,64,0}

\usepackage{booktabs} 
\definecolor{groupgray}{RGB}{235, 235, 235}
\definecolor{highlightpurple}{RGB}{235, 228, 248}
\usepackage{pifont}   
\usepackage{xcolor}   
\usepackage{array}    
\usepackage{makecell}
\usepackage{longtable}  
\usepackage{multirow}   
\usepackage{geometry}   
\newlength{\fullpagewidth}
\tcbuselibrary{breakable} 
\usepackage{listings}  
\usepackage{xcolor}  
\usepackage{etoc}
\usepackage{amsmath}
\usepackage{amssymb}
\usepackage{bbm}
\definecolor{lightred}{RGB}{255, 160, 160}
\definecolor{darkgreen}{RGB}{0, 120, 0}
\definecolor{darkblue}{RGB}{0, 70, 130}
\usepackage{geometry}    
\newcommand{\algcmt}[1]{\textcolor{teal}{\scriptsize // #1}}

\tcbuselibrary{skins, breakable}
\tcbset{
  graystyle/.style={
    enhanced, breakable,
    colback=gray!10,
    colframe=gray!60,
    colbacktitle=gray!70!black,
    coltitle=white,
    arc=2pt, boxrule=0.8pt,
    fonttitle=\bfseries,
    left=10pt, right=10pt, top=8pt, bottom=8pt,
  },
  orangestyle/.style={
    enhanced, breakable,
    colback=yellow!10,
    colframe=orange!70,
    colbacktitle=orange!80!yellow,
    coltitle=white,
    arc=3pt, boxrule=1.2pt,
    fonttitle=\bfseries,
    attach boxed title to top left={yshift=0pt, xshift=0pt},
    boxed title style={arc=3pt, boxrule=0pt,
      left=6pt, right=6pt, top=4pt, bottom=4pt},
  },
  bluestyle/.style={
    enhanced, breakable,
    colback=blue!5,
    colframe=blue!50,
    colbacktitle=blue!60,
    coltitle=white,
    arc=4pt, boxrule=1pt,
    fonttitle=\bfseries,
  },
}
\definecolor{mygreen}{RGB}{200, 230, 200}  
\definecolor{myblue}{RGB}{180, 220, 240}   
\definecolor{rulecolor}{RGB}{180, 180, 180}

\usepackage[ruled,vlined,linesnumbered]{algorithm2e}
\SetKwComment{TriComment}{$\triangleright$\ }{}
\newcommand{\methodname}{RecToolBench}
\DontPrintSemicolon

\usepackage{pifont}
\usepackage[table,xcdraw]{xcolor}
\usepackage{booktabs}

\title{\includegraphics[width=0.12\textwidth]{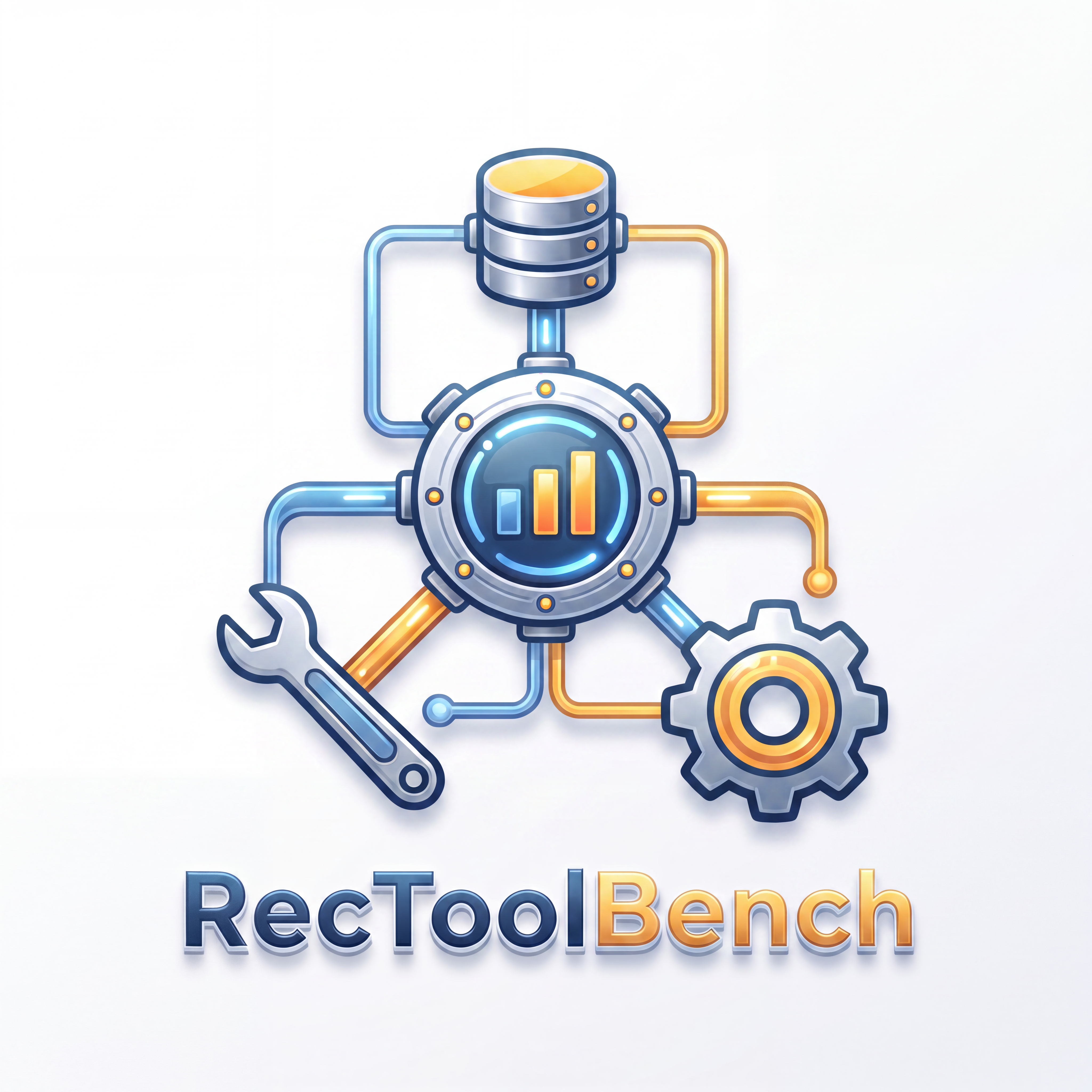} RecToolBench: Benchmarking Recommendation-Specific Tool Orchestration under Fuzzy User Intent}

\author{
\textbf{Xiao Chen}$^{1\dagger}$%
\thanks{Equal contribution.  $\dagger$Project leader. 
$\ddagger$Corresponding authors.},
\textbf{Yicheng Zhao}$^{2*}$,
\textbf{Yingying Wu}$^{2}$,
\textbf{Zhendong Chu}$^{3\ddagger}$\\
\textbf{Changyi Ma}$^{2\ddagger}$,
\textbf{Qingsong Wen}$^{3}$,
\textbf{Xuan Song}$^{2}$\\
$^{1}$The Hong Kong Polytechnic University \quad
$^{2}$Jilin University \quad
$^{3}$Squirrel AI Learning\\
\texttt{\{zc9uy@virginia.edu,changyima@jlu.edu.cn\}}
}

\begin{document}

\maketitle

\begin{abstract}

Recent advances in agentic recommender systems are shifting recommender systems from passive filtering engines to instruction-following agents that use external tools to resolve user intent. However, existing benchmarks often assume explicit user intent, simplified tool environments, or isolated function calls, leaving realistic tool orchestration for recommendation underexplored.
To bridge this gap, we propose \textbf{RecToolBench}, a Model Context Protocol (MCP)-based benchmark for evaluating tool-using recommender agents under fuzzy user instructions. RecToolBench contains more than 1,200 executable tasks across three recommendation domains, 13 MCP servers, and 32 tools, spanning single-tool calls, parallel tool calls, sequential tool chains, and hybrid tool orchestration.
We construct RecToolBench with a scalable \emph{synthesize--fuzzify--judge} pipeline that generates executable fuzzy recommendation tasks, and evaluates agent trajectories using rule-based execution checks and rubric-based LLM evaluation.
Experiments on representative LLMs show that syntactically valid tool calls do not guarantee successful recommendations. Models struggle with semantic parameter grounding, multi-step evidence integration, and grounded final recommendations, especially as orchestration complexity increases. Our results identify tool orchestration under fuzzy user intent as a major bottleneck for agentic recommender systems.
Our data and code are available at \url{https://github.com/ShawnChenn/RecToolBench}.

\end{abstract}

\etocdepthtag.toc{mainmatter}
\section{Introduction}

Agentic capabilities are reshaping recommender systems (RS) from \textit{passive} filtering engines into \textit{proactive}, instruction-following agents~\cite{huang2025survey,maragheh2025future}. A key paradigm in this transition is \textit{tool-using recommender agents}~\cite{zhao2024let,tang2026interactive}, which equip LLMs with external tools to iteratively clarify user intent and ground recommendations in retrieved evidence. Such capabilities are particularly well-suited to practical settings such as personalized e-commerce~\cite{zhang2026recthinker,chen2025c2kd}, where user requests are often open-ended and highly context-dependent.

Despite this progress, existing benchmarks remain limited in evaluating recommender agents under realistic tool-orchestration scenarios. General tool-use benchmarks typically assume explicit, well-specified instructions~\cite{yao2024tau,zhou2024webarena}, while recommendation-oriented benchmarks often focus on narrow settings (e.g., e-commerce QA) or isolated function calls~\cite{jin2024shopping,chen2025chineseecomqa}. As shown in Table~\ref{tab:tool_bench_compare}, these works largely overlook a common real-world challenge in recommendations: user instructions are often \textbf{fuzzy} and \textbf{underspecified}. In such cases, agents must infer latent intent, coordinate appropriate tools, and compose tool outputs into grounded recommendations, which means that tool orchestration in recommendations requires more than syntactically valid tool calls.

\begin{table}[t]
\centering
\small
\resizebox{1.0\linewidth}{!}{
\setlength{\tabcolsep}{2pt}
\renewcommand{\arraystretch}{1.2}
\begin{tabular}{lcccc}
\toprule
\textbf{Benchmark} &   \# \textbf{Tools} & \textbf{\makecell{End-to-end\\Performance}}  & \textbf{\makecell{Fuzzy \\ Instruction}}  & \textbf{\makecell{Graded \\ Difficulty}} \\
\midrule
{Shopping MMLU~\cite{jin2024shopping}}         & - & \textcolor{red}{\texttimes} & \textcolor{red}{\texttimes} & \textcolor{red}{\texttimes} \\
{ChineseEcomQA~\cite{chen2025chineseecomqa}}        & -   & \textcolor{red}{\texttimes} & \textcolor{red}{\texttimes} & \textcolor{red}{\texttimes}  \\
{$\tau$-bench~\cite{yao2024tau}}     & 28  & \textcolor{green}{\checkmark} & \textcolor{red}{\texttimes} & \textcolor{red}{\texttimes} \\
{ShoppingBench~\cite{wang2026shoppingbench}}     & 6   & \textcolor{green}{\checkmark} & \textcolor{red}{\texttimes} & \textcolor{orange}{$\circ$}\\
{Ours}   &  32 & \textcolor{green}{\checkmark} & \textcolor{green}{\checkmark} & \textcolor{green}{\checkmark} \\
\bottomrule
\end{tabular}}
\caption{Comparison with existing agent benchmarks that include recommendation scenarios. \textcolor{green}{\checkmark}: Fully addressed; \textcolor{orange}{$\circ$}: Partially addressed; \textcolor{red}{\texttimes}: Not addressed.}
\label{tab:tool_bench_compare}
\end{table}

\begin{figure*}[t]
    \centering
    \includegraphics[width=0.98\linewidth]{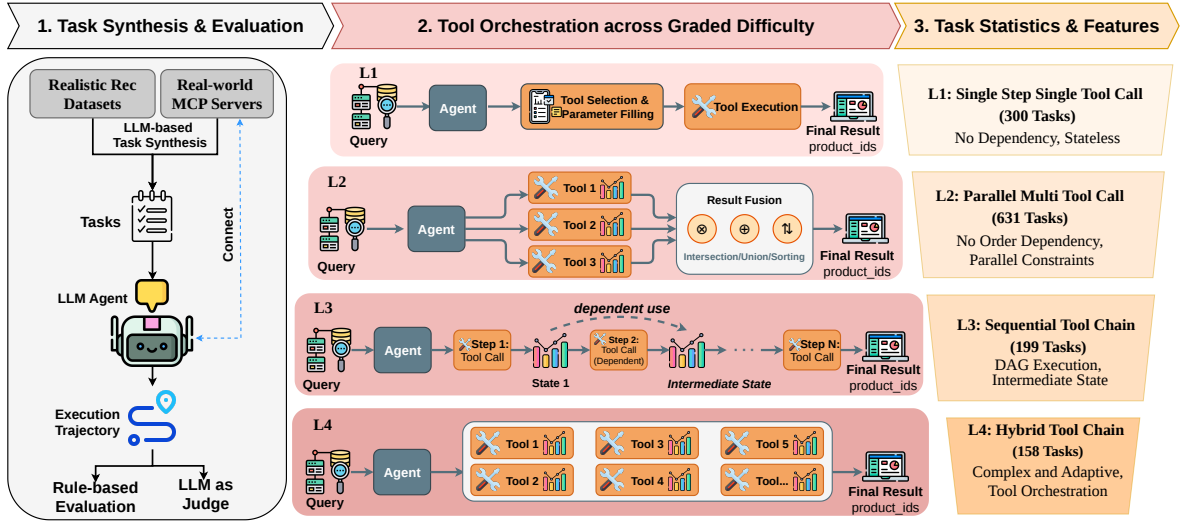}
    \caption{The overall data synthesis and agent evaluation pipeline, featuring a progressive four-level difficulty scheme. Tailored to address inherently fuzzy user instructions, the task complexity escalates from easy to hard: \textbf{L1} requires a single tool call; \textbf{L2} involves parallel tool invocations; \textbf{L3} demands serial dependency chains; and \textbf{L4} entails hybrid dependencies and tool orchestration.}
    \label{fig:data_frm}
\end{figure*}

Building a realistic benchmark for tool-using recommender agents requires addressing three key gaps. The first is \textit{fuzzy intent resolution}. Real user instructions rarely specify the structured parameters (e.g., category, budget, time window) that downstream tools require, forcing agents to infer latent intent rather than execute pre-formed queries. The second is \textit{compositional tool orchestration}. Recommendation tasks typically require parallel, serial, or cross-server tool dependencies, where intermediate outputs must be semantically propagated across steps, going well beyond isolated function calls. The third is the \textit{syntactic–semantic evaluation gap}. A successfully executed tool call does not imply correct orchestration, since agents often produce syntactically valid invocations whose parameters are semantically misaligned with the user's fuzzy intent, making execution-only metrics systematically over-optimistic.

To bridge these gaps, we propose \textbf{\methodname}, the first benchmark dedicated to evaluating tool orchestration in recommender agents under fuzzy user instructions. Built on the \emph{Model Context Protocol} (MCP)~\cite{hou2025model}, \methodname{} provides standardized tool invocation schema across servers. Using a scalable LLM-driven ``synthesize-fuzzify-judge'' pipeline, we construct over 1,200 executable tasks spanning three recommendation domains and 13 MCP servers. Each server encapsulates a cohesive suite of complementary tools (32 in total). As illustrated in Figure~\ref{fig:data_frm}, tasks are organized into a progressive four-level difficulty scheme that increases intent ambiguity and compositional dependency depth, ranging from single-tool calls to hybrid dependencies and orchestration.

Moreover, we conduct a comprehensive empirical study on Small Language Models (SLMs) and proprietary models. Each task is executed through single- or multi-turn MCP interactions and evaluated using rule-based checks for execution correctness and task completion, together with rubric-based LLM-as-a-Judge scoring for grounding and planning quality.
Our evaluation reveals several key findings.
First, \textit{SLMs struggle with correct tool use}: in L1, SLMs often produce valid calls but fail to select the correct tool or parameters.
Second, \textit{orchestration failures become more severe as task complexity increases}: in L2 and L3, models all struggle with parameter grounding, parallel decomposition, and dependency tracking, even when their schema compliance remains high.
Third, \textit{hybrid tool orchestration exposes a persistent execution--outcome gap}: in L4, models achieve strong rule-based scores, but their Hit Ratios and grounding scores remain much lower.
These findings show that agentic recommendation is bottlenecked by semantic parameter grounding, multi-step evidence integration, and grounded final recommendation.

\begin{figure*}[htbp]
\centering

\begin{subfigure}[b]{0.4\linewidth}
    \centering
    \includegraphics[width=\linewidth]{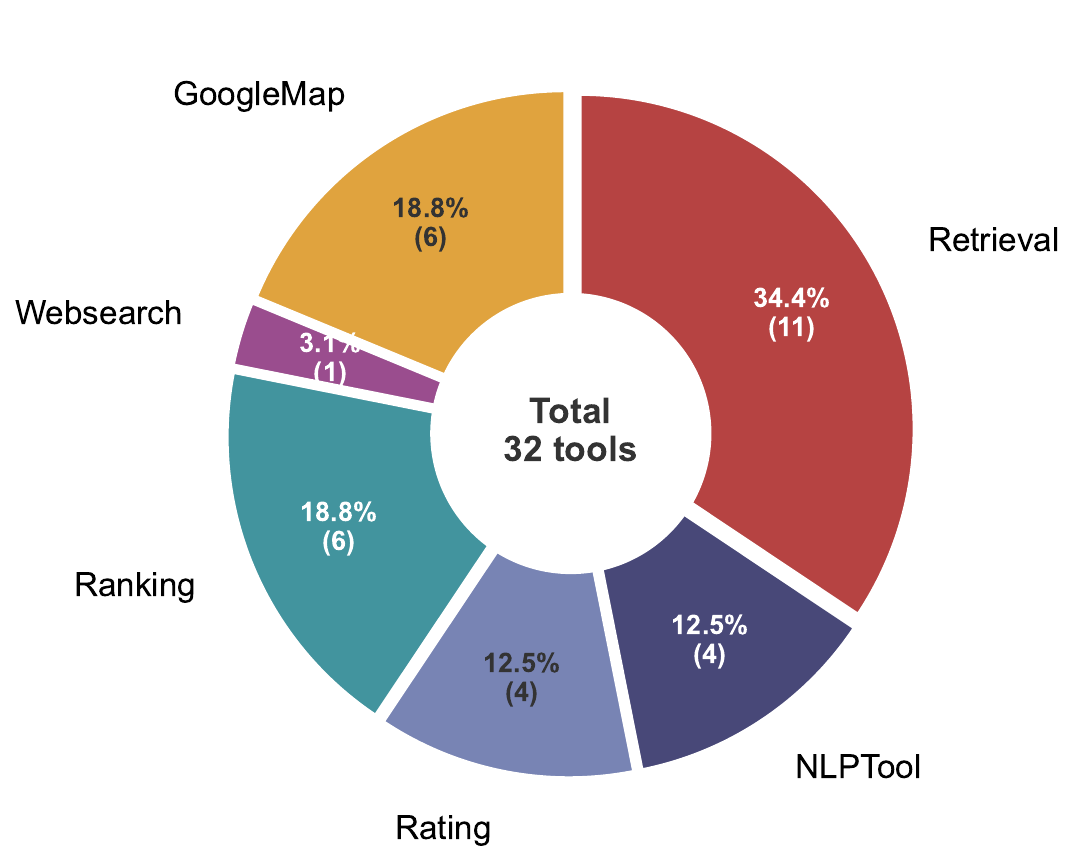}
    \caption{Category distribution of MCP servers.}
    \label{fig:bench-overview:a}
\end{subfigure}
\hfill
\begin{subfigure}[b]{0.52\linewidth}
    \centering
    \includegraphics[width=\linewidth]{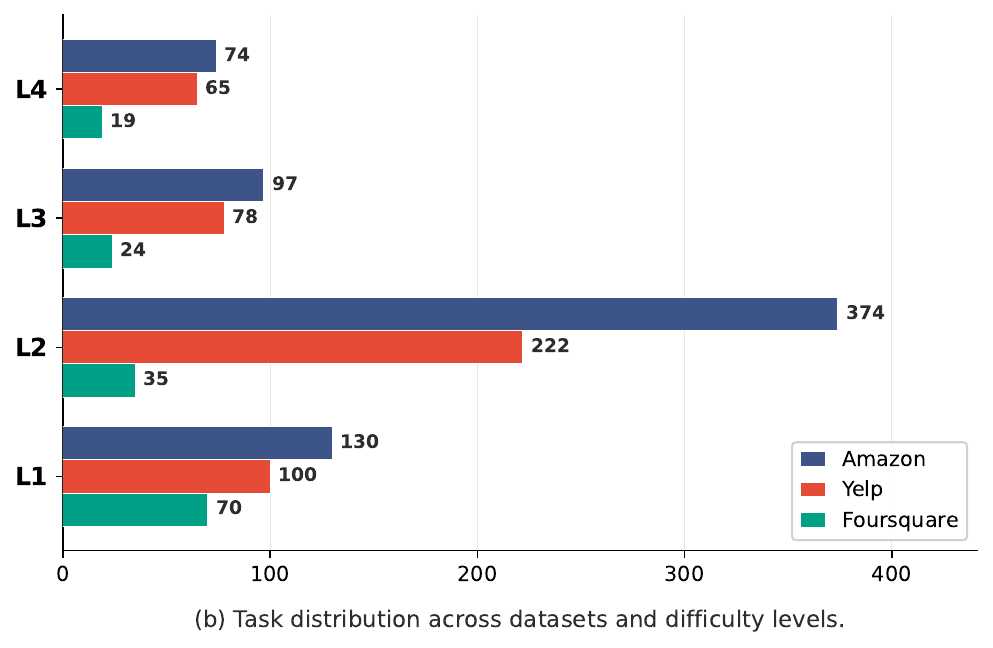}
    \caption{Task distribution across datasets and difficulty levels.}
    \label{fig:bench-overview:b}
\end{subfigure}
\caption{Overview of MCP server and task distribution in the \methodname{}.}
\label{fig:bench-overview}
\end{figure*}

Our contributions are summarized as follows:
\begin{itemize}
    \item We introduce the \textbf{first MCP-based benchmark for fuzzy recommendation tool orchestration}, covering diverse workflows, compositional dependencies, and graded difficulty levels.

    \item We develop a \textbf{scalable task synthesis pipeline} that generates executable fuzzy recommendation tasks grounded in real tool schemas, enabling agent trajectories to be evaluated through rule-based metrics and rubric-based LLM judging.

   \item We provide a \textbf{comprehensive empirical study} of SLMs and proprietary LLMs on these tasks. Our analysis shows that agent failures evolve with task complexity, from basic tool and parameter selection errors in SLMs to broader weaknesses in information grounding and dependency-aware orchestration in multi-step and hybrid tasks.

\end{itemize}

\section{Benchmark Construction}
\label{sec:task_synthesis}

\subsection{Problem Formulation}
We formalize each benchmark task as a tool-augmented partially observable Markov decision process (POMDP)~\cite{yao2024tau}. 
Given $n$ MCP servers $\Sigma=\{\sigma_1,\ldots,\sigma_n\}$, the available tool set is $\mathcal{T}=\bigcup_{j=1}^{n}\mathcal{T}_j$, where $\mathcal{T}_j$ is the set of tools exposed by server $\sigma_j$. 
Each task is represented as $\tau=(q,\mathcal{T},\mathcal{C},\mathcal{G})$, where $q$ is an underspecified user instruction, $\mathcal{C}=\{c_i\}_{i=1}^{N}$ is the candidate set, and $\mathcal{G}$ serves as the ground truth for both recommendation results and tool use. 
Specifically, $\mathcal{G}=(c^\star,\mathcal{A}^\star)$, where $c^\star\in\mathcal{C}$ is the target item and $\mathcal{A}^\star$ denotes the expected tool calls. 

The agent receives $(q,\mathcal{T})$ and interacts within a POMDP tuple $(\mathcal{S},\mathcal{A},\mathcal{O},\mathcal{P},\mathcal{R},\mathcal{U})$, denoting latent states $\mathcal{S}$, actions $\mathcal{A}$, observations $\mathcal{O}$, transition and observation models $\mathcal{P}$, rewards $\mathcal{R}$, and instructions $\mathcal{U}$. The action space consists of planning, tool invocation, and final recommendation: $\mathcal{A}=\mathcal{A}_{\mathrm{plan}}\cup\mathcal{A}_{\mathrm{tool}}\cup\mathcal{A}_{\mathrm{final}}$. A tool action is formatted as $a_{\mathrm{tool}}=\langle \sigma_j, t_k, \mathbf{p}_k\rangle$, representing the invocation of tool $t_k \in \mathcal{T}_j$ from server $\sigma_j \in \Sigma$ using arguments $\mathbf{p}_k$. The observation space $\mathcal{O}=\mathcal{O}_{\mathrm{tool}}\cup\mathcal{O}_{\mathrm{state}}$ combines external tool outputs ($\mathcal{O}_{\mathrm{tool}}$) and state feedback ($\mathcal{O}_{\mathrm{state}}$).
At step $t$, the agent selects an action $a_t\in\mathcal{A}$ conditioned on $q$, $\mathcal{T}$, and past observations $o_{<t}$. The environment subsequently yields a new observation according to $\mathcal{P}$, continuing until a final recommendation is generated or the interaction budget is exhausted.

\subsection{Task Synthesis}

A challenge in benchmark construction is generating realistic, naturally phrased tasks, which are grounded in executable tool-use workflows across MCP servers.
To this end, we ground task generation in user interaction sequences from three real-world datasets: Amazon Electronics\footnote{\url{http://jmcauley.ucsd.edu/data/amazon}}, Yelp 2019\footnote{\url{https://www.yelp.com/dataset/challenge}}, and Foursquare~\cite{noulas2011empirical}, corresponding to different recommendation scenarios.  

\paragraph{MCP Server Design.}
\methodname{} provides 13 domain-specific MCP servers across three recommendation domains, comprising 32 executable tools. These tools cover six core capabilities: retrieval, review and sentiment analysis, rating-based filtering, personalized recommendation, geo-spatial reasoning, and web search. Built on real-world recommendation datasets, they enable controlled yet realistic evaluation of tool-use recommender agents. Details of MCP servers are in Figure~\ref{fig:bench-overview:a} and Appendix~\ref{appen:mcp}.

\paragraph{Synthesis Pipeline.}
Accordingly, we propose \textit{AmbiFlow Synthesis}, an execution-aware and ambiguity-controlled pipeline for constructing realistic recommendation-oriented tool-use tasks. Given real-world user interaction sequences, AmbiFlow first identifies feasible tool-use workflows based on MCP tool schemas and user-item contexts. It then prompts a task synthesis LLM, i.e., GLM-5.1, with difficulty-level-specific requirements to generate explicit natural-language instructions along with the corresponding ground-truth tool sets. Next, AmbiFlow transforms each explicit instruction into a fuzzy one through a controlled fuzzification process, which is guided by five common ambiguity types: \textbf{underspecification}, \textbf{soft contradiction}, \textbf{vague quantifiers}, \textbf{preference uncertainty}, and \textbf{implicit constraints}. This process removes explicit tool names and procedural hints while introducing realistic uncertainty, yet preserves the underlying user intent. A final quality-filtering stage verifies the solvability, structural validity, and ambiguity reasonableness of each task. 
We further conduct human validation to assess instruction realism, workflow executability, and dependency correctness. We report these annotation details in Appendix~\ref{appendix:synthesis}

\begin{table*}[t]
  \centering
  \small
  \renewcommand{\arraystretch}{1.25}
  \setlength{\tabcolsep}{4pt}
  \resizebox{\textwidth}{!}{
  \begin{tabular}{
    >{\centering\arraybackslash}p{1.2cm}
    >{\centering\arraybackslash}p{6.8cm}
    >{\raggedright\arraybackslash}p{3cm}
    >{\centering\arraybackslash}p{2cm}
  }
  \toprule
  \textbf{Level} & \textbf{Fuzzy User Query} & \textbf{Tool-Use Workflow} & \textbf{Ambiguity Types} \\
  \midrule
  L1
    & ``I'm trying to find some camera stuff---maybe a hot shoe cover or bubble level for my Sony? I remember a brand called Foto\&Tech or something similar.''
    & \path|filter_items_by_attributes|
    & {\small Underspecification} \\
  \midrule
  L2 \newline (parallel)
    & ``I've got [candidate item asins] saved but can't remember which one I wanted. I think around 4 stars? Not super strict. Budget maybe under \$30 or \$40?''
    & \path|filter_products|
      \newline \path|filter_items_by_attributes|
      \newline \path|query_sentiment|
    & {\small Contradiction
      \newline \small Underspecification} \\
  \midrule
  L3 \newline (serial)
    & ``Find accessories for [user\_id]. Check their history first, I think it was a soundbar or speaker, maybe a keyboard? Get the details, then filter by that brand and price range.''
    & \path|get_user_history|
      \newline $\rightarrow$ \path|get_item_details|
      \newline $\rightarrow$ \path|filter_items_by_attributes|
      
    & {\small  Implicit Constraints
      \newline \small  Underspecification} \\
  \midrule
  L4 \newline (hybrid)
    & ``I bought [item asin] and want accessory recommendations, peripherals maybe, not 100\% sure. Search co-purchases and keywords at the same time. Check reviews and sentiment. Get details on top picks.''
    & \path|item2item_search| $\parallel$ \path|search_keyword|
       $\rightarrow$
       \path|search_reviews| $\parallel$ \path|query_sentiment|
       $\rightarrow$
       \path|get_item_details|
    & {\small Preference Uncertainty \newline
                   \small Underspecification \newline
                   \small Contradiction}   \\
  \bottomrule
  \end{tabular}}
  \caption{Representative task examples from each difficulty level (Amazon Electronics).
  Examples L2–L4 are drawn from a query set with relatively high ambiguity.}
  \label{tab:task_examples}
\end{table*}

\paragraph{Dataset Statistics}
The benchmark contains 1,288 synthesized tasks with grounded user intents and progressively increasing compositional complexity. Specifically, it includes 300 Level-1 tasks and 988 compositional tasks from Levels 2--4. Across datasets, the benchmark covers 675 Amazon, 465 Yelp, and 148 Foursquare tasks. Detailed statistics are provided in Fig~\ref{fig:bench-overview:b}. We also present fuzzy query examples in Table~\ref{tab:task_examples}.

\begin{algorithm}[t]
\small
\caption{Multi-turn Planning and Tool Execution for Benchmark Evaluation}
\label{alg:agent_rollout}
\SetAlgoNoLine
\DontPrintSemicolon
\SetAlgoNoEnd
\SetKwInput{KwIn}{\textbf{Input}}
\SetKwInput{KwOut}{\textbf{Output}}

\SetKwFunction{Plan}{Plan}
\SetKwFunction{Execute}{Execute}
\SetKwFunction{Slim}{Slim}
\SetKwFunction{UpdateBus}{UpdateBus}
\SetKwFunction{UpdateState}{UpdateState}
\SetKwFunction{Compress}{Compress}
\SetKwFunction{Synthesize}{Synthesize}

\KwIn{User instruction $u$, tool set $\mathcal{T}$, max rounds $T_{\max}$, token budget $B$}
\KwOut{Final recommendation $\hat{y}$, execution trajectory $H$}

\algcmt{Initialize $S_p$: planning state; $\mathcal{B}$: candidate bus; $H$: execution trajectory}

$S_p \leftarrow \emptyset$; $\mathcal{B} \leftarrow \emptyset$; $H \leftarrow \emptyset$\;

\For{$t=1$ \KwTo $T_{\max}$}{

    \algcmt{Build planner prompt}
    
    $z_t \leftarrow \textsc{RenderState}(S_p) \oplus \textsc{RenderBus}(\mathcal{B})$;

    \algcmt{Generate current turn's tool plan}
    
    $(\texttt{cont}_t, A_t) \leftarrow \pi_{\text{plan}}(u, z_t, \mathcal{T})$;

    \lIf{$\neg\,\texttt{cont}_t$ \textbf{or} $A_t = \emptyset$}{\textbf{break}}

    \algcmt{Parallel or serial tool execution}

    $O_t^{\text{raw}} \leftarrow \pi_{\text{exec}}(A_t)$\;

    \algcmt{Extract Item IDs from raw observation}

    $O_t \leftarrow \Slim(O_t^{\text{raw}})$\;

    \algcmt{Update candidate bus status}

    $\mathcal{B} \leftarrow \mathcal{B} \cup \textsc{ExtractBus}(O_t^{\text{raw}}, t)$\;
    
    \algcmt{Update agent internal states}

    $S_p \leftarrow \UpdateState(S_p,\; A_t,\; O_t)$\;

    \algcmt{LLM-based state compression}
    
    \If{$|S_p| > B$}{
        $S_p \leftarrow \Compress(S_p)$\;
    }

    $H \leftarrow H \cup \{(A_t, O_t)\}$\;
}
$\hat{y} \leftarrow \Synthesize(u,\; H)$\;
\algcmt{LLM synthesizes final answer from trajectory}

\Return{$\pi_{\text{final}}(u, H)$, $H$}\;

\end{algorithm}

\section{Evaluation Protocol and Metrics}

\subsection{Evaluation Protocol}

We evaluate agents using the multi-round plan--execute loop in Algorithm~\ref{alg:agent_rollout}. 
Each rollout maintains three complementary states: a compact planning state $S_p$ used in the planner prompt, a Candidate Bus $\mathcal{B}$ that stores candidate items and their metadata outside the prompt to reduce context load, and the full trajectory $H$ used for final synthesis and evaluation. 
At each round, the planner conditions on the user instruction, tool schemas, the compressed planning history, and the candidate summary maintained in $\mathcal{B}$. 
It then either predicts the next tool plan $A_t$ or terminates the rollout.

Tool outputs are stored in two views. Raw outputs $O_t^{\text{raw}}$ keep complete tool outputs, while planning outputs $O_t^{\text{plan}}$ retain only compact fields needed for subsequent planning, such as IDs. The Candidate Bus is updated from raw outputs to preserve reusable intermediate items without adding full payloads to the planner context. If the planner exceeds the context budget or returns an empty response, $S_p$ is compressed while preserving tool names, round order, key parameters, and critical findings. This design keeps long execution traces available for evaluation while maintaining a compact context for multi-round planning.
  
\begin{figure*}[t]
    \centering
    \includegraphics[width=\linewidth]{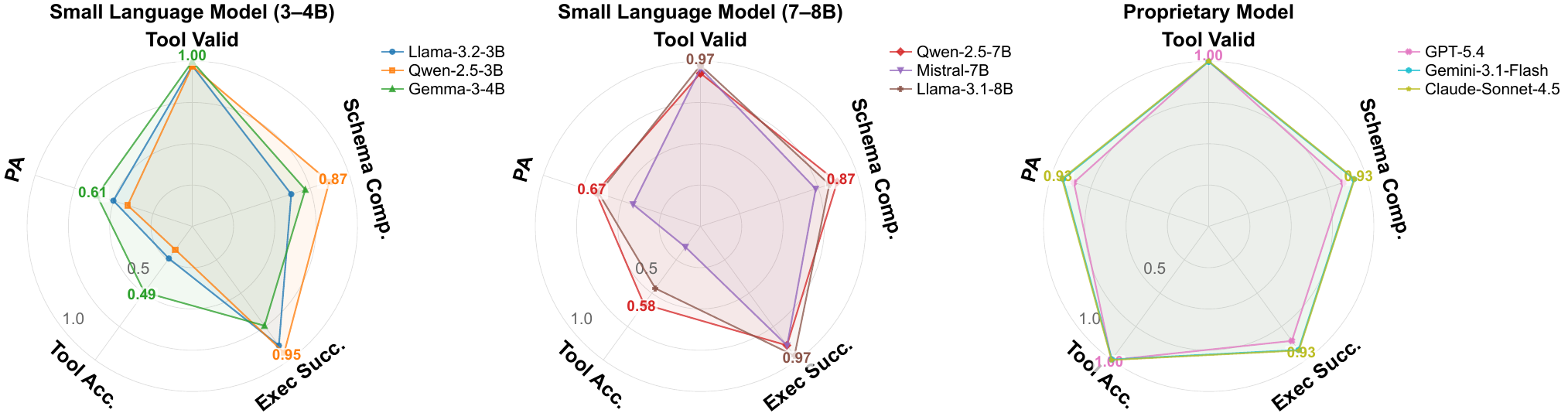}
    \caption{Evaluation results of representative LLM agents on atomic tool using abilities (L1).}
    \label{fig:L1_model_comparison_radar}
\end{figure*}

\subsection{Evaluation Metrics}
\label{sec:metrics}

Given an agent trajectory $\hat{\mathcal{G}}=\{(\hat{t}_j,\hat{\mathbf{p}}_j)\}_{j=1}^{n}$ and the available gold annotations, we evaluate tool use using two types of metrics. Rule-based evaluation focuses on the robustness of tool usage, while LLM-as-a-judge evaluation measures strategic quality with structured rubrics and prompt shuffling to mitigate positional bias.

\paragraph{Rule-based evaluation.}
We first check whether each executed call is valid and executable: \textit{Tool Name Validity} verifies that the predicted tool exists in the available tool set; \textit{Schema Compliance} checks whether its arguments satisfy the API schema; and \textit{Execution Success} requires the call to finish without error and return a non-empty result.

Then, we compare predicted tool usage with gold annotations. Single-tool tasks are evaluated using \textit{Tool Selection Accuracy}, while multi-tool tasks use \textit{Tool Selection F1} over predicted and gold tool sets. We further report orchestration and outcome metrics: \textit{Parallel Ratio} measures whether independent calls are issued early, \textit{Efficiency} penalizes redundant calls, and \textit{Hit Ratio} indicates whether the target item appears in the results.

\paragraph{LLM-as-a-Judge Evaluation.}
For L1 and L2, we report a reference-free \textit{Parameter Appropriateness (PA)} score, where an independent judge (DeepSeek-V3.2) assesses whether the generated parameters are semantically appropriate for the fuzzy instruction. For longer serial and hybrid trajectories (L3-L4), exact matching cannot fully capture execution quality, as agents may select plausible tools while violating dependencies, unnecessarily serializing independent calls, or producing unsupported recommendations. We therefore evaluate the full execution trace using an LLM judge. Serial workflows are assessed by \textit{Grounding} and \textit{Dependency Awareness}, while hybrid workflows are evaluated by \textit{Planning Coherence}, \textit{Multi-round Efficiency}, \textit{Parameter Appropriateness}, and \textit{Information Grounding}. Each dimension is scored from 1 to 10, averaged, and normalized to $[0,1]$. Detailed definitions and evaluation prompts are provided in Appendix~\ref{appendix:metrics}.

\paragraph{Remark.}
For L1--L3, we provide ground-truth workflows as evaluation references. \textbf{L1} contains a single annotated tool call. \textbf{L2} provides an unordered set of tools that can be invoked in parallel within one round. Thus, all tool-set metrics are permutation-invariant. \textbf{L3} provides the efficient sequential trajectory, explicitly capturing the required dependencies among tool calls. For \textbf{L4}, we do not provide a reference workflow because hybrid tasks may admit multiple valid combinations of parallel and sequential execution.

\begin{table*}[ht]
  \centering
  \renewcommand{\arraystretch}{1.1}
  \resizebox{\textwidth}{!}{
  \begin{tabular}{lccccccc}
    \toprule
    \multirow{2}{*}{\textbf{Model}}
      & \multicolumn{2}{c}{\textbf{Schema Understanding}}
      & \multicolumn{2}{c}{\textbf{Tool Usage}}
      & \multicolumn{2}{c}{\textbf{Planning}}
      & \multicolumn{1}{c}{\textbf{Completion}} \\
    \cmidrule(lr){2-3}
    \cmidrule(lr){4-5}
    \cmidrule(lr){6-7}
    \cmidrule(lr){8-8}
      & {\makecell{Schema \\ Compliance Rate}}
      & {\makecell{Execution \\ Success Rate}}
      & {\makecell{Tool Selection \\ F1 Score}}
      & {\makecell{Parameter \\ Appropriateness}}
      & {\makecell{Parallel \\ Rate}}
      & {\makecell{Efficiency \\ Score}}
      & {\makecell{Hit Ratio \\ @All}} \\
    \midrule

    \multicolumn{8}{c}{\cellcolor{groupgray}\textbf{Small Language Models}} \\

    Llama-3.2-3b-instruct
      & 97.77\% & 95.14\% & 41.35\% & 36.60\%
      & 33.57\% & 9.97\% & 17.54\% \\

    Qwen-2.5-3b-instruct
      & 93.96\% & 90.12\% & \cellcolor{myblue}49.20\% & 55.20\%
      & \cellcolor{myblue}42.20\% & \cellcolor{mygreen}45.38\% & 22.81\% \\

    Gemma-3-4b-instruct
      & 94.34\% & 88.83\% & 42.27\% & 35.70\%
      & 34.69\% & 23.49\% & 10.53\% \\

    Qwen-2.5-7b-instruct
      & \cellcolor{myblue}98.56\% & 96.45\%
      & \cellcolor{mygreen}52.66\% & 68.50\%
      & \cellcolor{mygreen}44.37\% & 30.17\% & 31.58\% \\

    Mistral-v0.3-7b-instruct
      & 90.09\% & 85.67\% & 45.52\% & 44.20\%
      & 40.39\% & 17.82\% & 17.54\% \\

    Llama-3.1-8b-instruct
      & \cellcolor{mygreen}99.79\% & 98.41\% & 44.16\% & 58.80\%
      & 40.33\% & 10.74\% & 28.07\% \\

    \midrule
    \multicolumn{8}{c}{\cellcolor{groupgray}\textbf{Proprietary Models}} \\

    GPT-5.4
      & 94.46\% & \cellcolor{mygreen}100.00\% & 43.55\%
      & \cellcolor{mygreen}93.00\%
      & 36.67\% & \cellcolor{myblue}34.07\% & 43.78\% \\

    Gemini-3.1-Flash-Lite
      & 86.66\% & \cellcolor{mygreen}100.00\% & 37.09\%
      & \cellcolor{myblue}85.90\%
      & 32.94\% & 16.81\% & \cellcolor{mygreen}55.59\% \\

    Claude-Sonnet-4.5
      & 96.58\% & \cellcolor{myblue}98.56\% & 42.42\%
      & 82.10\%
      & 39.37\% & 30.41\% & \cellcolor{myblue}47.29\% \\

    \bottomrule
  \end{tabular}}
  \caption{Performance Results with Synthesized User Instructions --- Parallel Tool Using Abilities (L2). The highest and second-highest scores are highlighted in
  \colorbox{mygreen}{\textcolor{black}{green}} and
  \colorbox{myblue}{\textcolor{black}{blue}}, respectively.}
  \label{tab:model_comparison2}
\end{table*}

\section{Experiment Results}
\noindent \textbf{Baselines}: We evaluate nine representative LLMs, including six small language models (SLMs)\footnote{Language Models with fewer than 10B parameters following~\cite{belcak2025small}}: Llama-3.2-3b-it, Qwen-2.5-3b-it, Gemma-3-4b-it, {Qwen-2.5-7b-it}, {Mistral-v0.3-7b-it}, and {Llama-3.1-8b-it}; and three proprietary models: {GPT-5.4}, {Gemini-3.1-Flash}, and {Claude-Sonnet-4.5}.

\noindent \textbf{Implementation Details}: To reduce order sensitivity in LLM-based judging~\cite{li2026evaluating}, we apply prompt shuffling and score averaging.
For each task, we randomly permute the evaluation axes and sub-dimensions within the scoring rubric, while keeping the rubric content unchanged. We run five shuffled evaluations per task, normalize each sub-dimension score to $[0,1]$, compute axis-level averages, and take the mean across runs as the final score.
Results are reported separately for each difficulty level (L1--L4), averaged over all tasks within that level, to isolate the effect of increasing orchestration complexity on model performance.

\subsection{Main Results}

\begin{table*}[t]
  \centering
  \renewcommand{\arraystretch}{1.1}
  \resizebox{\textwidth}{!}{
  \begin{tabular}{lcccccccc}
    \toprule
      & \multicolumn{5}{c}{\textbf{Rule-based}} 
      & \multicolumn{2}{c}{\textbf{Rubric-based}} 
      & \multicolumn{1}{c}{} \\
    \cmidrule(lr){2-6} \cmidrule(lr){7-8}
    \multirow{3}{*}{\textbf{Model}}
      & \multicolumn{2}{c}{\textbf{Schema Understanding}}
      & \multicolumn{1}{c}{\textbf{Tool Usage}}
      & \multicolumn{1}{c}{\textbf{Planning}}
      & \multicolumn{1}{c}{\textbf{Completion}}
      & \multicolumn{1}{c}{\textbf{Planning}} 
      & \multicolumn{1}{c}{\textbf{Grounding}}
      & \multirow{2}{*}{\textbf{\makecell{Overall \\ Score}}} \\
    \cmidrule(lr){2-3} \cmidrule(lr){4-4} \cmidrule(lr){5-5} \cmidrule(lr){6-6} \cmidrule(lr){7-8}
      & {\makecell{Schema \\ Compliance Rate}}
      & {\makecell{Execution \\ Success Rate}}
      & {\makecell{Tool Selection \\ F1 Score}}
      & {\makecell{Efficiency \\ Score}}
      & {\makecell{Hit \\ Ratio}}
      & {\makecell{Dependency \\ Awareness}}
      & {\makecell{Information \\ Grounding}} 
      & \\
    \midrule
    \multicolumn{9}{c}{\cellcolor{groupgray}\textbf{Small Language Models}} \\
    Llama-3.2-3b-instruct & \cellcolor{mygreen}97.08\% & 86.66\% & 44.00\%  & 67.24\% & 54.27\%  & 0.200 & 0.162 & 0.551 \\
    Qwen-2.5-3b-instruct & 96.62\% & 82.08\% & 53.34\%  & 67.44\% & 52.76\% & 0.253 & 0.286 & 0.580 \\
    Gemma-3-4b-instruct & 89.20\% & 35.95\% & 37.18\%  & 41.75\% & 21.11\% & 0.341 & 0.473 & 0.438 \\
    Qwen-2.5-7b-instruct & 96.96\% & 87.96\% & \cellcolor{mygreen}60.18\%  & 69.75\% & 52.76\% & 0.309 & 0.293 & 0.611 \\
    Mistral-v0.3-7b-instruct & 92.69\% & 77.41\% & 53.07\%  & 86.47\% & 65.83\% & 0.203 & 0.207 & 0.595 \\
    Llama-3.1-8b-instruct & \cellcolor{myblue}97.01\% & 92.15\% & \cellcolor{myblue}57.14\%  & 75.37\% & 59.30\% & 0.268 & 0.224 & 0.615 \\
    \midrule
    \multicolumn{9}{c}{\cellcolor{groupgray}\textbf{Proprietary Models}} \\
    GPT-5.4 & 93.65\% & 98.47\% & 50.43\% & \cellcolor{myblue}93.67\% & \cellcolor{mygreen}96.67\% & \cellcolor{mygreen}0.458 & \cellcolor{mygreen}0.615 & \cellcolor{mygreen}0.772 \\
    Gemini-3.1-Flash-Lite & 88.27\% & \cellcolor{myblue}99.82\% & 44.77\%  & 73.15\%  & 93.33\% & 0.275 & 0.332 &  0.657 \\
    Claude-Sonnet-4.5 & 91.01\% & \cellcolor{mygreen}99.95\% & 52.10\% & \cellcolor{mygreen}98.03\% & \cellcolor{myblue}95.00\% & \cellcolor{myblue}0.437 & \cellcolor{myblue}0.548 & \cellcolor{myblue}0.764 \\
    \bottomrule
  \end{tabular}}
  \caption{Performance Results with Synthesized User Instructions --- Sequential Tool Using Abilities (L3). The highest and second-highest scores are highlighted in 
\colorbox{mygreen}{\textcolor{black}{green}} and 
\colorbox{myblue}{\textcolor{black}{blue}}.
  }
  \label{tab:model_comparison3}
\end{table*}
\subsubsection{Atomic Tool Using Abilities (L1)}

We first report the L1 single-tool evaluation results in Figure~\ref{fig:L1_model_comparison_radar}. We have three key observations:

\noindent\textit{I. Proprietary models perform better in functional tool use.}
Proprietary models achieve 100\% Tool Selection Accuracy, with near-perfect schema compliance and execution success. 
In contrast, the best SLM, Qwen-2.5-7B-it, reaches 58\% Tool Selection Accuracy, despite maintaining high format compliance. 
This gap indicates that current SLMs remain unreliable for tool use, even in the simplest single-tool setting.

\noindent\textit{II. Format validity does not imply semantic correctness.}
SLMs often produce structurally valid tool calls, while failing at functional correctness. For example, Gemma-3-4B obtains 99\% Tool Name Valid Rate but only 49\% Tool Selection Accuracy, while Qwen-2.5-3B achieves 95\% Execution Success Rate but low Parameter Appropriateness. This indicates that format compliance alone is insufficient for semantically correct tool use.

\noindent\textit{III. Scaling up model size alone is not sufficient.}
Within the Qwen-2.5 family, the 7B variant significantly outperforms the 3B variant in Parameter Appropriateness. However, cross-family comparisons reveal that this relationship is not monotonic: Llama-3.1-8B and Gemma-3-4B achieve comparable Tool Selection Accuracy (46\% vs.\ 49\%). This suggests that instruction-tuning quality and tool-use alignment also play a critical role in recommendation-oriented tool use.

\noindent\textit{IV. Reference-free PA reveals weakness in parameter grounding.}
To mitigate the single-reference concern for parameter evaluation, we report reference-free \textit{Parameter Appropriateness (PA)} on L1, judged by DeepSeek-V3.2 from the fuzzy instruction. Figure~\ref{fig:L1_model_comparison_radar} shows that even when tool selection is correct, weaker models still exhibit substantially lower PA.

\subsubsection{Parallel Tool-Use Abilities (L2)}

We report the L2 parallel tool-use results in Table~\ref{tab:model_comparison2}. The results reveal several findings:

\noindent\textit{I. SLMs improve on schema-level execution but still lag in task completion.}
In L2, several SLMs are competitive in schema compliance and parallel planning, yet proprietary models achieve higher Hit Ratio. 
The gap mainly comes from tool-call effectiveness: Table~\ref{tab:avg_calls_rounds} shows that some SLMs issue as many, or even more, calls per task than proprietary models, but many of these calls are repetitive or use incorrect parameter values, often yielding empty results. 
This also explains the divergence between Schema Compliance and Execution Success Rate. 
Proprietary models can achieve higher execution success than schema compliance because MCP servers often tolerate minor deviations. 

\noindent\textit{II. Tool Selection F1 reflects different tool-use strategies.}
Proprietary models invoke more unique tools than the ground-truth plan, often adding auxiliary tools (e.g., \texttt{get\_item\_details}) to gather richer evidence. 
This lowers Tool Selection F1 but improves final recommendation result synthesis. 
In contrast, most SLMs call fewer tools and stay closer to the expected tool set, yielding higher F1 but lower task completion.

\noindent\textit{III. Parameter grounding is the central bottleneck under fuzzy intent (confirmed by reference-free PA).}
All SLMs struggle to ground correct arguments from fuzzy user queries when multiple calls must be coordinated in parallel. To mitigate the generator--evaluator alignment and single-reference concerns, we evaluate reference-free \textbf{Parameter Appropriateness (PA)} on L2 using DeepSeek-V3.2 as an independent judge (no access to the generated reference). The result indicates that parameter grounding remains a robust bottleneck across evaluation protocols.

\noindent\textit{IV. High parallelism does not necessarily imply high efficiency.}
Several SLMs achieve high Parallel Rate yet low Efficiency Score, indicating that their parallel calls are frequently redundant rather than strategically decomposed. While efficient parallel tool use requires coordinated, task-relevant calls, rather than merely increasing the number of simultaneous tool invocations.

\begin{table*}[t]
  \centering
  \renewcommand{\arraystretch}{1.2}
  \resizebox{\textwidth}{!}{
  \begin{tabular}{lccccccccc}
    \toprule
      & \multicolumn{4}{c}{\textbf{Rule-based}} 
      & \multicolumn{4}{c}{\textbf{Rubric-based}} 
      & \multicolumn{1}{c}{} \\
    \cmidrule(lr){2-5} \cmidrule(lr){6-9}
    \multirow{3}{*}{\textbf{Model}}
      & \multicolumn{2}{c}{\textbf{Schema Understanding}}
      & \multicolumn{1}{c}{\textbf{Tool Usage}}
      & \multicolumn{1}{c}{\textbf{Completion}}
      & \multicolumn{2}{c}{\textbf{Planning}} 
      & \multicolumn{2}{c}{\textbf{Tool Grounding}}
      & \multirow{2}{*}{\textbf{\makecell{Overall \\ Score}}} \\
    \cmidrule(lr){2-3} \cmidrule(lr){4-4} \cmidrule(lr){5-5} \cmidrule(lr){6-7} \cmidrule(lr){8-9}
      & {\makecell{Schema \\ Compliance Rate}}
      & {\makecell{Execution \\ Success Rate}}
      & {\makecell{Tool Selection \\ F1 Score}}
      & {\makecell{Hit \\ Ratio}}
      & {\makecell{Planning \\ Coherence}} 
      & {\makecell{Multi-round \\ Efficiency}} 
      & {\makecell{Parameter \\ Appropriateness}} 
      & {\makecell{Information \\ Grounding}}
      & \\
    \midrule
    \multicolumn{10}{c}{\cellcolor{groupgray}\textbf{Small Language Models}} \\
    Llama-3.2-3b-instruct & 92.82\% & 85.61\% & 59.06\%  & 20.27\% & 0.178 & 0.130 & 0.231 & 0.161 & 0.410\\
    Qwen-2.5-3b-instruct & 92.57\% & 88.75\% & 71.10\%  & 20.27\% & 0.235 & 0.234 & 0.291 & 0.154 & 0.455\\
    Gemma-3-4b-instruct & 96.91\% & 90.06\% & 64.97\%  & 20.27\% & 0.167 & 0.127 & 0.212 & 0.145 & 0.422\\
    Qwen-2.5-7b-instruct & 97.27\% & 89.59\% & \cellcolor{mygreen}81.39\% & 25.68\% & 0.268 & 0.212 & 0.321 & 0.193 & 0.492\\
    Mistral-v0.3-7b-instruct & 87.18\% & 72.53\% & 68.03\%  & 24.32\% &  0.182 & 0.142 & 0.181 & 0.153 & 0.397\\
    Llama-3.1-8b-instruct & \cellcolor{myblue}99.14\% & 91.81\% & 67.40\% & 28.38\% & 0.207 & 0.132 & 0.276 & 0.207 & 0.461\\
    \midrule
    \multicolumn{10}{c}{\cellcolor{groupgray}\textbf{Proprietary Models}} \\
    GPT-5.4 & \cellcolor{mygreen}100.00\% & \cellcolor{myblue}99.61\% & \cellcolor{myblue}78.42\% & \cellcolor{mygreen}54.05\% & \cellcolor{mygreen}0.408 & \cellcolor{mygreen}0.370 & \cellcolor{mygreen}0.500 & \cellcolor{mygreen}0.418 & \cellcolor{mygreen}0.627\\
    Gemini-3.1-Flash-Lite & \cellcolor{mygreen}100.00\% & \cellcolor{mygreen}100.00\% & 73.52\% & 39.19\% & 0.347 & 0.277 & 0.392 & \cellcolor{myblue}0.384 & 0.566\\
    Claude-Sonnet-4.5 & \cellcolor{mygreen}100.00\% & \cellcolor{mygreen}100.00\% & 78.15\% & \cellcolor{myblue}44.59\% & \cellcolor{myblue}0.355 & \cellcolor{myblue}0.349 & \cellcolor{myblue}0.431 & 0.368 & \cellcolor{myblue}0.591\\
    \bottomrule
  \end{tabular}}
  \caption{Performance Results with Synthesized User Instructions --- Hybrid Tool Using Abilities (L4). The highest and second-highest scores are highlighted in 
\colorbox{mygreen}{\textcolor{black}{green}} and 
\colorbox{myblue}{\textcolor{black}{blue}}.
  }
  \label{tab:model_comparison4}
\end{table*}
\subsubsection{Sequential Tool-Use Abilities (L3)}
We report the results on sequential tool chain tasks in Table~\ref{tab:model_comparison3}. Key observations are as follows:

\noindent\textit{I. Sequential dependencies expose a clear gap between schema-level execution and true task completion.}
Several SLMs still achieve high Schema Compliance and Execution Success in L3. For example, Llama-3.1-8B reaches 97.01\% Schema Compliance and 92.15\% Execution Success. However, compared with proprietary models, SLMs remain weaker in Hit Ratio and rubric-based metrics, especially Dependency Awareness. This suggests that the main challenge in L3 is not producing syntactically valid tool calls, but correctly carrying intermediate information across dependent steps.

\noindent\textit{II. High tool selection quality does not guarantee successful sequential orchestration.}
Qwen-2.5-7B achieves the highest Tool Selection F1 among all SLMs at 60.18\%. However, its Hit Ratio is only 52.76\%, far below proprietary models. This gap indicates that identifying the right set of tools is not sufficient in L3: the agent must also preserve dependency structure, ground arguments from prior observations, and execute calls in a coherent multi-step workflow.

\noindent\textit{III. Proprietary models maintain high completion through stronger dependency tracking and grounding.}
GPT-5.4 and Claude-Sonnet-4.5 sustain very high Hit Ratios of 96.67\% and 95.00\%, respectively. Their advantage is matched by much stronger rubric-based scores, including Dependency Awareness and Information Grounding. This pattern suggests that proprietary models are better at tracking intermediate states and grounding later tool calls on earlier results.

\noindent\textit{IV. SLMs exhibit heterogeneous failure modes under sequential dependencies.}
Llama-3.1-8B achieves the highest overall score among SLMs. Mistral-v0.3-7B leads open models in Hit Ratio (65.83\%) and Planning Efficiency, but remains weak in Dependency Awareness and Information Grounding. Gemma-3-4B shows relatively strong Information Grounding but low Hit Ratio (21.11\%). This suggests that robust L3 performance requires multiple coupled abilities rather than isolated strengths.

\begin{table}[t]
    \centering
    \small
    \resizebox{\linewidth}{!}{
    \begin{tabular}{l cc cc cc}
    \toprule
    \multirow{2}{*}{\textbf{Model}}
      & \multicolumn{2}{c}{\textbf{L2}}
      & \multicolumn{2}{c}{\textbf{L3}}
      & \multicolumn{2}{c}{\textbf{L4}} \\
    \cmidrule(lr){2-3} \cmidrule(lr){4-5} \cmidrule(lr){6-7}
      & Calls & Rounds & Calls & Rounds & Calls & Rounds \\
    \midrule
    \rowcolor{groupgray}
    \textbf{Ground Truth (average)} & \textbf{3.3} & \textbf{1.0} & \textbf{3.0} & \textbf{3.0} & - & - \\
    \midrule
    \multicolumn{7}{c}{\cellcolor{groupgray}\textbf{Small Language Models}} \\
    Llama-3.2-3B     & 37.7 & 4.8 & 6.1  & 6.0 & 41.7 & 5.4 \\
    Qwen-2.5-3B      & 10.0 & 2.8 & 5.9  & 5.8 & 15.5 & 2.7 \\
    Gemma-3-4B       & 16.5 & 4.6 & 6.2  & 6.0 & 27.0 & 5.6 \\
    Qwen-2.5-7B      & 12.9 & 4.4 & 5.8  & 5.8 & 16.5 & 4.6 \\
    Mistral-7B       & 20.0 & 4.5 & 7.5  & 5.9 & 22.7 & 5.3 \\
    Llama-3.1-8B     & 32.5 & 5.0 & 6.1  & 6.0 & 68.2 & 6.0 \\
    \midrule
    \multicolumn{7}{c}{\cellcolor{groupgray}\textbf{Proprietary Models}} \\
    GPT-5.4           & 9.8  & 2.1 & 11.7 & 2.9 & 11.8 & 3.1 \\
    Claude-Sonnet-4.5 & 14.7 & 3.6 & 11.2 & 3.3 & 11.9 & 4.0 \\
    Gemini-3.1-Flash-Lite    & 18.8 & 4.4 & 23.7 & 5.3 & 19.0 & 4.3 \\
    \bottomrule
    \end{tabular}}
    \caption{Average tool calls and rounds across models.}
    \label{tab:avg_calls_rounds}
  \end{table}

\subsubsection{Hybrid Tool-Use Abilities (L4)}
We report the results on complex hybrid recommendation tasks in Table~\ref{tab:model_comparison4}. The results reveal several critical observations.

\noindent\textit{I. L4 hybrid tasks widen the gap between tool execution and recommendation success.}
By combining structural complexity with semantic ambiguity, L4 requires agents to coordinate multi-round tool use while resolving underspecified intent. In this setting, task completion drops sharply: the best proprietary models reach only 54.05\% and 44.59\% Hit Ratio, while SLMs remain at 20\%--28\%. Thus, long-horizon hybrid recommendation remains difficult even when tool execution is largely reliable.

\noindent\textit{II. High rule-based scores do not ensure hybrid-task success.}
In L4, many models achieve high Schema Compliance, Execution Success, and Tool Selection F1, but their Hit Ratio remains substantially lower.
This gap suggests that L4 task requires coherent multi-round coordination and evidence-grounded recommendation.

\noindent\textit{III. Hybrid tool-use and grounding remains challenging.}
Proprietary models generally achieve higher overall performance, suggesting stronger capabilities in planning and evidence grounding.
SLMs show competitive results on several rule-based metrics, but their performance is less consistent on rubric-based dimensions and final recommendation accuracy.
This discrepancy indicates that hybrid-task success requires more than correct tool invocation; models must also integrate evidence across multiple steps and maintain grounding throughout the reasoning process.

\subsection{Round and Tool Calls Across Models}
Table~\ref{tab:avg_calls_rounds} shows that proprietary models adhere most closely to the ground-truth execution budget, maintaining round counts that match the expected task structure despite utilizing more tool calls. In contrast, Llama models exhibit severe redundancy, often repeating identical calls rather than contributing new information. SLMs demonstrate a different inefficiency at L3, largely proceeding serially by issuing one call per round until the limit, instead of parallelizing independent steps. Collectively, these patterns indicate that efficient tool use hinges on redundancy control, dependency-aware parallelization, and reliable termination, rather than merely on the ability to invoke tools.

\subsection{Error analysis}
We further inspect GPT-5.4 trajectories in L4. 
The two dominant failure modes are dependency reasoning failures (54.1\%) and target retrieval misses (53.6\%), showing that strong proprietary models still struggle to maintain cross-step dependencies and propagate intermediate evidence into the final recommendation.

\section{Related Work}

\subsection{Tool-Augmented Recommender Systems}

Recent work augments LLM-based recommender systems with external tools. InteRecAgent~\cite{huang2025recommender} and ToolRec~\cite{zhao2024let} explore tool orchestration and surrogate-user modeling, while RecGPT-V2~\cite{yi2025recgpt} and i$^2$Agent~\cite{xu2025iagent} extend tool use to multi-agent collaboration and safety enforcement. However, these systems differ in their tool sets, task formulations, and evaluation protocols, motivating a unified benchmark for systematic evaluation and fine-grained diagnosis of agents' tool-calling capabilities in recommendation scenarios~\cite{zhang2025recommendation,fan2022comprehensive,chen2023fairly}. Additional related work and system details are provided in the Appendix.



\subsection{Related Benchmarks}

Existing related benchmarks fall into two categories. \textbf{(i) Recommendation-specific benchmarks without tool calling} evaluate LLM knowledge and reasoning in e-commerce scenarios. Shopping MMLU~\cite{jin2024shopping} and ChineseEcomQA~\cite{chen2025chineseecomqa} use static QA formats that reduce recommendation to isolated text generation, while WebShop~\cite{yao2022webshop} and WebArena~\cite{zhou2024webarena} enable interactive evaluation but focus on product-search and purchase tasks rather than broader recommendation workflows.
\textbf{(ii) Recommendation-specific benchmarks with tool calling} are most closely related to our work but remain confined to narrow e-commerce settings. ShoppingBench~\cite{wang2026shoppingbench} provides a sandbox with structured APIs, yet it offers no explicit difficulty hierarchy for diagnosing workflows with varying complexity.
In contrast, our benchmark is \textit{recommendation-native} and defines \textit{explicit difficulty levels} for fine-grained evaluation.
\section{Conclusion}
We introduce \textbf{\methodname{}}, an MCP-based benchmark for evaluating tool-using agents under fuzzy recommendation instructions. It provides executable tasks with graded difficulty and diverse orchestration patterns, from single-tool use to hybrid workflows. Results show that current models often execute tools correctly but remain weak in parameter grounding, dependency tracking, evidence integration, and grounded recommendation, making complex tool orchestration a key bottleneck for reliable agentic recommender systems.

\section*{Limitations}
Despite the expanded evaluation across L1--L4, this work still has several limitations. First, \methodname{} is evaluated in a controlled MCP environment, which supports reproducible execution but does not fully reflect the noise and latency of real-world industrial-level recommendation systems. Second, the current benchmark covers only a limited subset of recommendation scenarios, tool ecosystems, and user behaviors, given the diversity of industrial production settings. Third, while we now include both small language models and proprietary frontier models, our study still does not cover the full spectrum of agent designs due to cost constraints. Expanding along these dimensions is an important direction for future work.

\clearpage

\bibliography{custom}
\clearpage

\appendix
\etocdepthtag.toc{appendix}     
\etocsettagdepth{mainmatter}{none}
\etocsettagdepth{appendix}{subsection}
\etocsettocstyle{\section*{Appendix}}{}
\tableofcontents   

\section{Author Contributions}

Xiao led the project, coordinated its overall development and execution, and took primary responsibility for drafting the manuscript. He designed and contributed to L1--L4 data synthesis, data verification, and analysis of the experimental results. YC contributed to L1--L2 data synthesis and partial experiments. YY contributed to L3 data synthesis and partial experiments. 
Zhendong provided valuable suggestions and guidance on idea initiation and experimental design throughout the project.
CY contributed to figure refinement.  
QS and XS provided valuable support throughout the project.

\section{Detailed Related Work}

\subsection{Tool-Augmented Recommender Systems}
\label{app:tool_augmented_rs}

The tool-use capabilities of LLMs have been extensively studied. Representative benchmarks include $\tau$-Bench~\cite{yao2024tau} and BFCL~\cite{patil2025bfcl} for multi-turn tool-use evaluation, as well as ToolHop~\cite{ye2025toolhop} for evaluating complex tool-dependent reasoning.

Building on these advances, integrating external tools into LLM-based recommender systems has emerged as an active research direction. InteRecAgent~\cite{huang2025recommender} adopts an ``LLM-as-brain, RS-as-hands'' architecture in which the LLM coordinates recommendation tools. ToolRec~\cite{zhao2024let} introduces surrogate-user modeling and attribute-conditioned tool chains. RecGPT-V2~\cite{yi2025recgpt} scales to production via a multi-agent framework with specialized agents collaborating through tool-calling protocols. i$^2$Agent~\cite{xu2025iagent} further positions the agent as a protective intermediary that enforces safety constraints via external tools.

\subsection{Related Benchmark}
More broadly, general tool-calling benchmarks~\cite{yu2026benchmarking,zhong2025complexfuncbench,li2025tool} emphasize multi-step function composition, and MCP-based benchmarks~\cite{yan2025mcpworld,gao2025mcp,liu2025mcpeval} focus on standardized interaction protocols. Neither line captures the fuzzy intent grounding, item-centric retrieval, and heterogeneous tool compositions that characterize real-world recommendations. 

\section{Details of Used MCP Servers}
\label{appen:mcp}
In Table \ref{tab:amazon_functions}, Table \ref{tab:foursquare_functions} and Table \ref{tab:yelp_functions}, we show the detailed descriptions for the involved MCP servers and the associated tools for the three individual tasks based on Amazon Electronics, Foursquare and Yelp data.

\onecolumn

\newcommand{\mcpcolA}{0.19\fullpagewidth} 
\newcommand{\mcpcolB}{0.14\fullpagewidth} 
\newcommand{\mcpcolC}{0.19\fullpagewidth} 
\newcommand{\mcpcolD}{0.36\fullpagewidth} 
\setlength{\tabcolsep}{3pt}
\renewcommand{\arraystretch}{1.15}

\begin{longtable}{|p{\mcpcolA}|p{\mcpcolB}|p{\mcpcolC}|p{\mcpcolD}|}
    \hline
    \textbf{Server Name} & \textbf{Tool Name} & \textbf{Input} & \textbf{Output} \\
    \hline
    \endfirsthead
    
    \multicolumn{4}{|c|}{{\tablename\ \thetable{} —— Continued from previous page}} \\
    \hline
    \textbf{Server Name} & \textbf{Tool Name} & \textbf{Input} & \textbf{Output} \\
    \hline
    \endhead
    
    \hline
    \multicolumn{4}{|r|}{{Continued on next page}} \\
    \endfoot
    
    \endlastfoot
    
    \multirow{9}{\mcpcolA}{Retrieval MCP Server}
        & Get Item Details & Item ID (has fall back to fuzzy item title search) & JSON of full item metadata \\
        \cline{2-4}
        & \multirow{4}{\mcpcolB}{Filter Items by Attributes} & Category & \multirow{4}{\mcpcolD}{JSON of filtered items} \\
            \cline{3-3}
            & & brand & \\
            \cline{3-3}
            & & min price & \\
            \cline{3-3}
            & & max price & \\
        \cline{2-4}
        & \multirow{2}{\mcpcolB}{Item2Item Search}
            & Query Text (try match against item title)
                & \multirow{2}{\mcpcolD}{JSON of referenced items based on similarity score} \\ 
            \cline{3-3}
            & & Top-K & \\
        \cline{2-4}
        & \multirow{2}{\mcpcolB}{Search Keywords}
            & Query (try match against item description)
                & \multirow{2}{\mcpcolD}{JSON of filtered items by keywords, sort by hit rate} \\
            \cline{3-3}
            & & Top-K & \\
    \hline
    \multirow{3}{\mcpcolA}{NLPTool MCP Server}
        & Query Sentiment & Min Positive Rate & JSON string with items matching the sentiment threshold \\
        \cline{2-4}
        & \multirow{2}{\mcpcolB}{Search Reviews}
            & Keywords (List of keywords to search for in review)
                & \multirow{2}{\mcpcolD}{JSON string with items whose reviews mention the specified aspects} \\
            \cline{3-3}
            & & Search In & \\
    \hline
    \multirow{3}{\mcpcolA}{Recommendation MCP Server}
        & \multirow{2}{\mcpcolB}{Rank Items}
            & User ID
                & \multirow{2}{\mcpcolD}{JSON of ranked items according to user id, contains id, score and rank} \\ 
            \cline{3-3}
            & & Item IDs & \\
        \cline{2-4}
        & Get User History & User ID & JSON with user id, histroy length, last 20 items \\
    \hline
    \multirow{8}{\mcpcolA}{Rating MCP Server}
        & \multirow{7}{\mcpcolB}{Filter Products}
            & Min Rating
                & \multirow{7}{\mcpcolD}{JSON string with products matching the rating criteria, sorted by rating descending} \\ 
            \cline{3-3}
            & & Max Rating & \\
            \cline{3-3}
            & & Min Reviews & \\
            \cline{3-3}
            & & Max Reviews & \\
            \cline{3-3}
            & & Category & \\
            \cline{3-3}
            & & Sort By & \\
            \cline{3-3}
            & & Limit (maximum number of returns, default 20) & \\
        \cline{2-4}
        & Compare Ratings & Item IDs (list) & JSON string with rating comparison for the specified items \\
    \hline
    \multirow{2}{\mcpcolA}{Websearch MCP Server}
    & \multirow{2}{\mcpcolB}{Web Search} & Query & \multirow{2}{\mcpcolD}{JSON string with search results including title, link, snippet, and optional knowledge graph} \\
        \cline{3-3}
        & & Num Results (default 10) & \\
    \hline
    \caption{Details of tools and descriptions in used Amazon Electronics MCP Servers.}
    \label{tab:amazon_functions}
\end{longtable}

\vspace{5pt}

\begin{longtable}{|p{\mcpcolA}|p{\mcpcolB}|p{\mcpcolC}|p{\mcpcolD}|}
    \hline
    \textbf{Server Name} & \textbf{Tool Name} & \textbf{Input} & \textbf{Output} \\
    \hline
    \endfirsthead
    
    \multicolumn{4}{|c|}{{\tablename\ \thetable{} —— Continued from previous page}} \\
    \hline
    \textbf{Server Name} & \textbf{Tool Name} & \textbf{Input} & \textbf{Output} \\
    \hline
    \endhead
    
    \hline
    \multicolumn{4}{|r|}{{Continued on next page}} \\
    \endfoot
    
    \endlastfoot
    
    \multirow{8}{\mcpcolA}{POI Retrieval MCP Foursquare Server}
        & Get POI Details & Venue ID & JSON of venue item \\
        \cline{2-4}
        & \multirow{4}{\mcpcolB}{Search POIs}
            & Query (try match against category name)
                & \multirow{4}{\mcpcolD}{JSON of venues that met requirements (query like category name, or location within distance)} \\ 
            \cline{3-3}
            & & Location (format: [lat, lng]) & \\
            \cline{3-3}
            & & Distance (in meters) & \\
            \cline{3-3}
            & & Top-K & \\
        \cline{2-4}
        & \multirow{3}{\mcpcolB}{Filter POIs}
            & Category
                & \multirow{3}{\mcpcolD}{JSON of venues that exact match category name or location or min checkins} \\
            \cline{3-3}
            & & Min Checkins & \\
            \cline{3-3}
            & & Location & \\
    \hline
    \multirow{3}{\mcpcolA}{Recommendation MCP Foursquare Server}
        & \multirow{2}{\mcpcolB}{Rank Items}
            & User ID
                & \multirow{2}{\mcpcolD}{JSON of ranked venues by user ID; contains venue ID, score, and rank} \\
            \cline{3-3}
            & & Item IDs & \\
        \cline{2-4}
        & Get User History & User ID & JSON with user id, histroy length, last 20 items \\
    \hline
    \multirow{5}{\mcpcolA}{GoogleMap MCP Server}
        & \multirow{2}{\mcpcolB}{Google Geo}
            & Address
                & \multirow{2}{\mcpcolD}{JSON of geo object} \\

            \cline{3-3}
            & & City & \\
        \cline{2-4}
        & Google ReGeo & LatLng & JSON of geo object \\
        \cline{2-4}
        & \multirow{2}{\mcpcolB}{Filter POIs}
            & Origin (LatLng)
                & \multirow{2}{\mcpcolD}{JSON of geo object with distance and travel time} \\

            \cline{3-3}
            & & Destination (LatLng) & \\
    \hline
    \caption{Details of tools and descriptions in used Foursquare MCP Servers.}
    \label{tab:foursquare_functions}
\end{longtable}

\begin{longtable}{|p{\mcpcolA}|p{\mcpcolB}|p{\mcpcolC}|p{\mcpcolD}|}
    \hline
    \textbf{Server Name} & \textbf{Tool Name} & \textbf{Input} & \textbf{Output} \\
    \hline
    \endfirsthead
    
    \multicolumn{4}{|c|}{{\tablename\ \thetable{} —— Continued from previous page}} \\
    \hline
    \textbf{Server Name} & \textbf{Tool Name} & \textbf{Input} & \textbf{Output} \\
    \hline
    \endhead
    
    \hline
    \multicolumn{4}{|r|}{{Continued on next page}} \\
    \endfoot
    
    \endlastfoot
    
    \multirow{9}{\mcpcolA}{Retrieval MCP Yelp Server}
        & Get Business Details & Query Text (of business name) & JSON of business item \\
        \cline{2-4}
        & \multirow{3}{\mcpcolB}{Search Businesses} & Query (against name, address, categories) & \multirow{3}{\mcpcolD}{JSON of businesses that met either query or around location, only top k} \\
            \cline{3-3}
            & & Location & \\
            \cline{3-3}
            & & Top-K & \\
        \cline{2-4}
        & \multirow{3}{\mcpcolB}{Filter Businesses} & Categories (exact category, list flatten) & \multirow{3}{\mcpcolD}{JSON of businesses that met exact criteria} \\

            \cline{3-3}
            & & Location & \\
            \cline{3-3}
            & & Open Timestamp (UTC Timestamp where business is open) & \\
        \cline{2-4}
        & Recall Similar Items & Query Text (business name) & JSON of business frequently visited together \\
    \hline
    \multirow{3}{\mcpcolA}{NLPTool MCP Server}
        & Query Sentiment & Min Positive Rate & JSON string with items matching the sentiment threshold \\
        \cline{2-4}
        & \multirow{2}{\mcpcolB}{Search Reviews}
            & Keywords (List of keywords to search for in review)
                & \multirow{2}{\mcpcolD}{JSON string with items whose reviews mention the specified aspects} \\

            \cline{3-3}
            & & Search In & \\
    \hline
    \multirow{3}{\mcpcolA}{Recommendation MCP Server}
        & \multirow{2}{\mcpcolB}{Rank Items}
            & User ID
                & \multirow{2}{\mcpcolD}{JSON of ranked items by user ID; contains item ID, score, and rank} \\

            \cline{3-3}
            & & Item IDs & \\
        \cline{2-4}
        & Get User History & User ID & JSON with user id, histroy length, last 20 items \\
    \hline
    \multirow{11}{\mcpcolA}{Rating MCP Server}
        & \multirow{10}{\mcpcolB}{Filter Products}
            & Min Rating
                & \multirow{9}{\mcpcolD}{JSON string with products matching the rating criteria, sorted by rating descending} \\

            \cline{3-3}
            & & Max Rating & \\
            \cline{3-3}
            & & Min Reviews & \\
            \cline{3-3}
            & & Max Reviews & \\
            \cline{3-3}
            & & Category & \\
            \cline{3-3}
            & & Business IDs (list) & \\
            \cline{3-3}
            & & Item IDs (list) & \\
            \cline{3-3}
            & & Sort By & \\
            \cline{3-3}
            & & Limit (maximum number of returns, default 20) & \\
        \cline{2-4}
        & Compare Ratings & Item IDs (list) & JSON string with rating comparison for the specified items \\
    \hline
    \multirow{5}{\mcpcolA}{GoogleMap MCP Server}
        & \multirow{2}{\mcpcolB}{Google Geo}
            & Address
                & \multirow{2}{\mcpcolD}{JSON of geo object} \\

            \cline{3-3}
            & & City & \\
        \cline{2-4}
        & Google ReGeo & LatLng & JSON of geo object \\
        \cline{2-4}
        & \multirow{2}{\mcpcolB}{Filter POIs}
            & Origin (LatLng)
                & \multirow{2}{\mcpcolD}{JSON of geo object with distance and travel time} \\

            \cline{3-3}
            & & Destination (LatLng) & \\
    \hline
    \caption{Details of tools and descriptions in used Yelp MCP Servers.}
    \label{tab:yelp_functions}
\end{longtable}

\twocolumn

\section{Benchmark Details}
\label{sec:data_detail}

  \subsection{Synthesis Pipeline Details}
  \label{appendix:synthesis}

  \paragraph{Ambiguity Fuzzification.}
  We introduce ambiguity from five categories: \textit{underspecification}, \textit{soft contradiction}, \textit{vague quantifiers}, \textit{preference uncertainty}, and
  \textit{implicit constraints}. Each task is assigned one of three intensity levels---low, moderate, and high---sampled with a 40\%/40\%/20\% distribution.

  \paragraph{Quality Filtering Criteria.}
  Filtering criteria are tailored to each level: Level 1 verifies schema executability and parameter validity; Level 2 additionally checks parallel independence among
  required tools; Level 3 validates dependency-chain correctness; Level 4 verifies the hybrid parallel-serial structure and fan-in dependencies. 

\paragraph{Human Validation of Query Quality.}

Synthesized query tasks from each domain and difficulty level are manually inspected by two annotators with experience in recommender systems and tool use.
Annotators evaluate each instance along four dimensions: (i)~whether the fuzzy instruction is natural and realistic, (ii)~whether the intended recommendation goal is recoverable, (iii)~whether the ground-truth tool workflow is executable under the MCP tool schemas, and (iv)~whether dependency-chain annotations are correct for multi-step tasks.
The two annotators show high agreement across these validation dimensions, with Cohen's $\kappa$ of 0.78.
Disagreements are resolved through discussion, and low-quality instances are removed or regenerated.

\begin{table*}[ht]
\centering
\scriptsize
\setlength{\tabcolsep}{4pt}
\renewcommand{\arraystretch}{1.12}
\begin{tabular}{>{\raggedright\arraybackslash}p{3.8cm}|>{\raggedright\arraybackslash}p{10.8cm}}
\hline
\textbf{Servers \& Tools} & \textbf{Task Description} \\
\hline
\textbf{Amazon L2} \newline
\textbf{Servers:} NLP Tool, Rating, Retrieval \newline
\textbf{Useful Tools (5):} compare\_ratings, search\_keyword, filter\_items\_by\_attributes, search\_reviews, item2item\_search
&
Hey, so I've been looking at a few electronics items lately — got my eye on these specific ones: [\textit{candidate item list}]. I'm kind of overwhelmed trying to figure out which one to actually go with.\par
Basically I want something that's decent quality and won't drain my wallet — ideally under \$100 for something in the electronics/accessories realm. I've seen some good reviews on a couple of these but haven't really dug into the details yet.\par
Oh, and I was also curious about what people typically pair with the EVGA Power Booster — maybe that could help me decide?\par
If you could take a look at these items and give me your thoughts on which one seems like the best pick, I'd really appreciate it. I'm not in a huge rush or anything, just trying to make a smart choice before I buy.
\\
\hline
\textbf{Yelp L3} \newline
\textbf{Servers:} Google Maps, Yelp NLP Tool, Yelp Rating, Yelp Rec, Yelp Retrieval \newline
\textbf{Useful Tools (5):} google\_geo, search\_businesses, google\_distance, google\_regeo, recall\_similar\_items
&
I'm looking for some dinner options near my location at 123 Main Street in San Francisco. I'm pretty hungry for pizza tonight and want to stay within walking distance—maybe around 2 kilometers or so.\par
First, could you find pizza places in that area? Once you have those results, I'd like to see the top 3 highest-rated ones. Then, can you check how far each one is from my address and filter out anything that's more than 2000 meters away?\par
If any pizza places are still in the running after that distance filter, I'd love to get their full addresses along with their ratings, review counts, and walking distances so I can make a decision.\par
Oh, and just in case there aren't any good pizza options within that walking range, I have a backup plan. There's a place called Devon Donut \& Bagel Company (business ID: qu0qoXXxVVDxMaof1SFzXg) that I really like. If needed, could you find me restaurants similar to that one from this list: [\textit{candidate item list}]? Then check which of those similar places are within about 3000 meters from me and rank them by how similar they are to that Devon place.
\\
\hline
\end{tabular}
\caption{Examples of tasks in \methodname{}, selected from tasks that require parallel tool invocation or multi-turn sequential tool use.}
\label{tab:benchmark_task_examples}
\end{table*}

\subsection{Prompts for Task Synthesis}
\label{app:prompts}
We detail the prompts used in our task synthesis process below, including user query instruction generation, prompts used for fuzzy matching and data quality filtering.

\section{Evaluation Metric for each Level}
\label{appendix:metrics}
\subsection{L1 Metrics}

In the Level 1 single-tool evaluation scenario, the task is relatively straightforward. We therefore use 4 \textbf{rule-based metrics} to assess the agent's \textcolor{lightred}{schema understanding} ability. Correct parameter specification, together with successful execution, indicates that the model has successfully resolved the recommendation task.

Here, let $i=(n, a)$ represent a single tool invocation produced by the agent for a given task, where $n$ is the tool name and $a$ is the argument object. Assume there are $N$ independent tool invocation tasks in total. Each rule-based metric is defined as follows:

\begin{enumerate}
    \item \textbf{Tool Name Valid Rate} ($T_\text{valid}$) calculates the fraction of tool invocations whose tool name belongs to the available tool set $\mathcal{T}_\text{available}$, denoted as
    $\ T_\text{valid}=\frac{1}{N}\sum_{j=1}^{N}\mathbf{1}\big[n_j\in\mathcal{T}_\text{available}\big]$.
     
    \item \textbf{Schema Compliance Rate} ($S_\text{acc}$) assesses the fraction of tool invocations whose arguments strictly conform to the selected tool's input schema (i.e., required fields and data types).
    
    \item \textbf{Execution Success Rate} ($R_\text{success}$) quantifies the success rate of executing the tool invocation without runtime errors and producing valid outputs, denoted as $
R_\text{success}=\frac{1}{N}\sum_{j=1}^{N}\mathbf{1}\big[\text{exec}(j)=\text{Success}\big]
$.

\item \textbf{Tool Selection Accuracy} ($T_\text{acc}$) measures the accuracy of tool choices compared to the ground-truth tool, denoted as
$T_\text{acc}=\frac{1}{N}\sum_{i=1}^{N}\mathbf{1}\big[n_i=\hat{n}_i\big]$,
where $\hat{n}_i$ is the ground-truth tool name for task $i$.

\item \textbf{Parameter Appropriateness} ($S_{\text{PA}}$) evaluates whether the generated arguments are semantically appropriate under fuzzy intent. We employ an independent LLM judge (DeepSeek-V3.2; different family from our synthesis model) to score appropriateness from the fuzzy user query, \emph{without access to the generated reference}. Scores are normalized to $[0,1]$.
\end{enumerate}

\subsection{L2 Metrics}
In the Level-2 Multi-Tool Parallel scenario, the task is ideally completed in a single round. Beyond schema understanding and tool usage, we also evaluate \textcolor{darkgreen}{planning effectiveness}, and end-to-end task completion:

\begin{enumerate}

    \item \textbf{Schema Understanding Ability} evaluates input schema adherence and execution validity. We adopt two metrics established in L1: \textit{Schema Compliance Rate} ($S_\text{acc}$) and \textit{Execution Success Rate} ($R_\text{success}$). In the L2 context, these metrics are aggregated across all tool invocations.
    
   \item \textbf{Tool Usage Quality} assesses the correctness of tool selection and parameter passing. \textit{Tool Selection F1 Score} ($T_\text{F1}$) evaluates the appropriateness of tool selection by balancing precision (accuracy) and recall (completeness).

    \item {\textbf{Planning Effectiveness}} evaluates the efficiency of the execution plan, specifically the agent's ability to minimize redundancy and exploit opportunities for parallel execution. 
    
    \textit{Parallel Ratio} ($R_{\text{para}}$) quantifies the alignment between the first-round invocations and the ground-truth plan: $R_{\text{para}} = \frac{|T_{\text{r1}} \cap T_{\text{exp}}|}{\max(|T_{\text{exp}}|, |T_{\text{r1}}|)}$. An optimal score ($R_{\text{para}} = 1.0$) indicates exact alignment.
    
    \textit{Efficiency Score} ($S_{\text{eff}}$) evaluates the execution efficiency relative to the ground-truth plan, defined as: $S_{\text{eff}} = \frac{N_{\text{exp}}}{\max(N_{\text{exp}}, N_{\text{total}})}$, where $N_{\text{total}}$ denotes the total number of tools actually invoked across all rounds. 
    The score reaches its optimum of $1.0$ when the agent invokes exactly the expected tools without redundancy, and it decreases when redundant calls occur.

    \item \textbf{Task Completion} measures the final recommendation quality. \textit{Hit Ratio} ($H_{\text{ratio}}$) evaluates whether the final recommended item matches the ground-truth target.
\end{enumerate}

\subsection{L3 Metrics} 
In Level 3, given the fuzzy and complex nature of the tasks, we employ a comprehensive evaluation framework combining rule-based metrics and LLM-as-judge scoring.

\begin{itemize}
    \item \textbf{Rule-based Evaluation}.  Let $E = \{e_1, \ldots, e_k\}$ be the set of all tool invocations during execution. Since we have the ground-truth set of required tools for Level 3 tasks, we evaluate the agent's multi-turn tool usage trajectory along the following dimensions:

    \begin{enumerate}
    
    \item \textbf{Schema Understanding Ability}: 
    We adopt Schema Compliance ($S_{\text{acc}}$): $S_{\text{acc}} = \frac{|\{e \in E : \text{valid\_schema}(e)\}|}{|E|}$, where $\text{valid\_schema}(e)$ indicates parameter adherence to the schema;
    and Execution Success ($R_{\text{suc}}$): $R_{\text{suc}} = \frac{|\{e \in E : \text{success}(e)\}|}{|E|}$, where $\text{success}(e)$ denotes error-free execution.

    \item \textbf{Tool Usage Quality}: We adopt {Tool Selection F1} ($T_{\text{F1}}$) from Level 2, calculated over the agent's entire trajectory $E$.

    \item \textbf{Planning Efficiency Score}: 
    We define sequential efficiency score ($S_{\text{eff}}$): 
    $S_{\text{eff}} = 1 - \frac{|\{e \in E : \text{is\_duplicate}(e)\}|}{|E|}$,
    where $\text{is\_duplicate}(e)$ indicates a redundant invocation with identical tool and arguments.

    \item \textbf{Task Completion}: We utilize the end-to-end metric \textit{Hit Ratio} ($R_{\text{hit}}$), measuring the alignment between the final result and the Ground Truth.
    
    \end{enumerate}
    
    \item \textbf{Rubric-based LLM Evaluation.} We employ DeepSeek-V3.2~\cite{liu2025deepseek} as LLM-as-a-Judge to evaluate two dimensions: \textbf{Information Grounding},
  assessing whether the final answer is derived from tool execution results rather than hallucinations, and \textbf{Dependency Awareness}, verifying whether the agent
  correctly follows the dependency chain by using outputs from previous steps. Each dimension is scored on a 1--10 scale using a fine-grained rubric, then normalized to    
  $[0, 1]$ for standardized comparison.
    
\end{itemize}

\subsection{L4 Metrics}   Level 4 evaluates hybrid workflows combining parallel and serial tool use. We combine rule-based metrics with rubric-based LLM-as-judge scoring for comprehensive evaluation.                                                                                                                
                       
\begin{itemize}
    \item \textbf{Rule-based Evaluation.} We evaluate three dimensions following the metric definitions in Level 3: (1) \textbf{Schema Understanding} ($S_{\text{acc}}$, $R_{\text{suc}}$), (2) \textbf{Tool Usage Quality} via Tool Selection F1 ($T_{\text{F1}}$), Unlike lower-level tasks, Level 4 hybrid workflows usually do not have a unique reference parameterization: downstream arguments may depend on the agent's intermediate results. Therefore, exact-match rule-based Parameter Accuracy is not appropriate, and we instead evaluated it semantically through LLM-as-a-judge. and
   (3) \textbf{Task Completion} via Hit Ratio ($R_{\text{hit}}$).

   \item \textbf{Rubric-based LLM Evaluation.} We employ DeepSeek-V3.2~\cite{liu2025deepseek} as the judge, given the task specification, execution trace, and final answer. L4 does not prescribe a unique execution sequence, but retains annotations of required tools. Four dimensions are
   scored on a 1--10 scale and normalized to $[0,1]$: \textbf{Planning Coherence} ($S_{\text{PC}}$), whether the agent correctly decomposes the task, parallelizes
  independent tools, respects serial dependencies, and performs fan-in aggregation, \textbf{Multi-round Efficiency} ($S_{\text{ME}}$), whether the agent avoids unnecessary 
  sequentialization, redundant calls, and excessive retries, \textbf{Parameter Appropriateness} ($S_{\text{PA}}$), whether parameters are semantically reasonable given task
   constraints, user intent, and upstream outputs, and \textbf{Information Grounding} ($S_{\text{IG}}$), whether the final response is faithfully grounded in tool outputs  
  without hallucinated attributes or unsupported conclusions.
\end{itemize}

\onecolumn

\begin{figure*}[t]
\centering
\noindent\textbf{Data Quality Filter}\par\smallskip
\begin{tcolorbox}[
    colframe=blue!50!black,
    colback=blue!5,
    sharp corners,
    boxrule=1pt,
    width=\textwidth,
    arc=4mm,
    fontupper=\small,
    title=Task Quality Assessment Prompt,
    breakable,
]
\textbf{System Role:}\\[-2pt]
You are an expert evaluator for AI agents that execute serial tool chains.
Score strictly based on evidence in the execution trace.
Be critical: default to 4--5 unless strong evidence supports higher.

\noindent\textbf{User:}\\[-1pt]
You must assign scores \textbf{only based on evidence} from the task, solution, and tool usage. Your evaluation should be:
\begin{itemize}[leftmargin=1.5em, itemsep=0pt, topsep=0pt, parsep=0pt, partopsep=0pt]
  \item Objective (avoid being influenced by language fluency or formatting)
  \item Justified (include specific reasons tied to each score)
  \item Robust against bias (ignore narrative style, verbosity, or formatting polish)
\end{itemize}
\vspace{-1pt}
\noindent You are evaluating an AI agent that completed a task by calling tools \textbf{one at a time} in a sequential chain, where each step may depend on results from the previous step.

\noindent\textbf{TASK DESCRIPTION}: ``\{task\}''\\[-2pt]
\textbf{EXECUTION TRACE}:\\[-5pt]
\{execution\_trace\}\\[-2pt]
\textbf{AGENT'S FINAL ANSWER}: ``\{final\_answer\}''

\vspace{1pt}\noindent\textcolor{rulecolor}{\rule{\linewidth}{0.4pt}}\vspace{1pt}

\noindent\textbf{Scoring Dimensions} — score each on a scale of 1--10.

\textbf{1. Grounding ($G$)}\\
Does the agent's final answer faithfully reflect what the tools actually returned? Penalize unsupported claims or hallucinations.
\noindent\small Score 9--10 if 90--100\% of factual claims are directly grounded in tool outputs; 7--8 if 70--80\% are grounded with minor unsupported details; 4--6 if 40--60\% are grounded with noticeable hallucination; and 1--3 if less than 30\% is grounded and the agent largely ignores tool results.

\textbf{2. Dependency Awareness ($D$)}\\
Does each step correctly use the output of the previous step? Penalize broken information flow.
\noindent\small Score 9--10 if 90--100\% of inter-step dependencies are correctly propagated; 7--8 if 70--80\% are correct with one broken link; 4--6 if 40--60\% are correct and the chain is partially broken; and 1--3 if less than 30\% are correct and the agent treats steps as independent.

\noindent Return \textbf{only} the following JSON:
\begin{small}
\begin{verbatim}
{
  "grounding_reasoning": "explain grounding...",
  "dependency_awareness_reasoning": "explain dependencies",
  "grounding": <int 1-10>, "dependency_awareness": <int 1-10>
}
\end{verbatim}
\end{small}
\end{tcolorbox}
\end{figure*}


\onecolumn

\noindent\textbf{Fuzzy Prompt}
\begin{tcolorbox}[
    orangestyle, title={Fuzzing Prompt}]
\noindent\textbf{Goal:} Convert ONE detailed multi-tool sequential task into a NATURAL,
CONVERSATIONAL user request.
The original task involves a CHAIN of tool calls where later tools depend
on earlier tools' outputs.

\medskip
\noindent\textbf{Input (detailed task):}\\
\{detailed\_task\}

\medskip
\noindent\textbf{Available tools:} \{len(tools)\} tools (but don't mention them in the fuzzy version)

\medskip
\noindent\textcolor{rulecolor}{\rule{\linewidth}{0.4pt}}

\medskip
\noindent\textbf{AMBIGUITY INJECTION:}\\
\{ambiguity\_config\}\\[4pt]
{[}Sampled per task from: AMBIGUITY LEVEL in \{LOW, MODERATE, HIGH\}\\
\phantom{[}with AMBIGUITY TYPES drawn from \{UNDERSPECIFICATION, SOFT\_CONTRADICTION,\\
\phantom{[}VAGUE\_QUANTIFIERS, PREFERENCE\_UNCERTAINTY, IMPLICIT\_CONSTRAINTS\}{]}

\medskip
\noindent\textcolor{rulecolor}{\rule{\linewidth}{0.4pt}}

\medskip
\noindent\textbf{CRITICAL RULES (Level3 sequential tasks):}
\begin{itemize}[leftmargin=1.5em, itemsep=2pt, topsep=2pt, parsep=0pt]
  \item The task naturally involves MULTIPLE STEPS in sequence.
  \item You MUST preserve the sequential/dependency nature of the request
        (e.g., ``find X first, then use X to get Y'').
  \item Use natural sequential connectors: ``first'', ``then'', ``after that'',
        ``once you have'', ``based on the results''.
  \item Do NOT flatten the task into a single parallel request.
  \item Apply the specified ambiguity level and types, but ALWAYS keep the
        sequential flow inferable.
\end{itemize}

\medskip
\noindent\textbf{STYLE:}
\begin{itemize}[leftmargin=1.5em, itemsep=2pt, topsep=2pt, parsep=0pt]
  \item Write like a real user describing a multi-step need.
  \item Use conversational language; do not sound like a benchmark task.
  \item Do NOT mention tool names or technical implementation details.
  \item Embed specific values naturally in conversation.
\end{itemize}

\medskip
\noindent\textbf{PRESERVE CRITICAL DATA:}
\begin{itemize}[leftmargin=1.5em, itemsep=2pt, topsep=2pt, parsep=0pt]
  \item Keep exact numeric thresholds (ratings, prices, counts) when present.
  \item Keep ALL user IDs, item IDs (ASINs), and entity IDs EXACTLY as they
        appear. Do NOT replace them with generic terms like ``my'' or ``the user''.
  \item Keep all candidate item ASINs/IDs that appear in the task --- these are
        INPUT data the agent needs, not target answers to hide.
  \item Use approximate language when appropriate: ``around'', ``about'',
        ``somewhere between''.
  \item ALWAYS USE relative time references (e.g., ``past few weeks'',
        ``recently'', ``in the next month''; not specific dates).
\end{itemize}

\medskip
\noindent\textbf{HARD CONSTRAINTS:}
\begin{itemize}[leftmargin=1.5em, itemsep=2pt, topsep=2pt, parsep=0pt]
  \item Never reveal the TARGET/ANSWER item identity directly (no exact target
        title, no target ASIN, no quoted seed target name).
  \item BUT you MUST keep all candidate item ASINs/IDs that appear in the
        task --- these are input data, not answers.
\end{itemize}

\medskip
\noindent Return ONLY the natural, conversational fuzzy description.
\end{tcolorbox}

\noindent\textbf{Data Quality Filter}
\begin{tcolorbox}[
    colframe=blue!50!black,
    colback=blue!5,
    sharp corners,
    boxrule=1pt,
    width=\linewidth,
    arc=4mm,
    fontupper=\small,
    title=Task Quality Assessment Prompt,
    breakable,
]
\begin{lstlisting}
Task Quality Assessment Prompt

Purpose: Evaluate task quality on solvability, utility, and consistency dimensions

Evaluate this task's quality on three dimensions:

Task Description:
{task_description}

Dependency Analysis (the serial chain logic and tool dependencies):
{dependency_analysis}

Fuzzy Description (what the agent sees):
{fuzzy_description}

Available Tools:
{tool_descriptions}

____________________________________________________________

EVALUATION CRITERIA:
1. SOLVABILITY (1-10):

10: All required data is provided, tools perfectly match needs, clear success criteria
8-9: Task is clearly solvable with the given tools, minor ambiguities acceptable
6-7: Mostly solvable but some steps may be challenging or unclear
4-5: Significant gaps in tool coverage or data requirements
1-3: Task cannot be meaningfully completed with available tools

Consider:

Are all necessary tools available?
Is all required data provided (no external dependencies)?
Can the agent achieve the stated goal with these tools based on the function and output of the tools?
Are success criteria clear and measurable?

2. UTILITY (1-10):

10: Perfectly simulates a real user recommendation scenario with rich personalization (budget, brand preference, feature requirements), multi-criteria decision-making (rating, review count, collaborative filtering), and clear practical value for shopping, dining, or location exploration
8-9: Strong recommendation scenario with meaningful user preferences and reasonable complexity; agent needs to synthesize signals from multiple tools to make justified recommendations
6-7: Basic recommendation task but lacks realistic personalization nuances or requires only simple single-tool queries; provides some value but could be more representative of real-world decision-making
4-5: Artificial or toy recommendation task; user intent is unrealistic, preferences are trivial or missing, and the task doesn't reflect actual shopping/exploration behavior
1-3: Completely contrived scenario with no resemblance to how users seek recommendations; fails to test agent's ability to balance multiple factors

Consider:

Does the task reflect a realistic user seeking recommendations (shopping, dining, places to visit)?
Are user preferences authentic (budget, brand affinity, functional needs, occasion context)?
Does the agent need to combine multiple signals (ratings, reviews, collaborative patterns, attributes) to make a well-justified recommendation?
Would the recommendations be genuinely useful for someone making a real decision?
Is the complexity aligned with typical user decision-making processes?


3. CONSISTENCY (1-10):

10: The fuzzy_description perfectly mirrors the dependency_analysis, same user intent, same key constraints (thresholds, categories, brands, keywords), same filtering/ranking logic, and the user's natural language request would naturally lead an agent to execute exactly the serial chain described
8-9: The fuzzy_description captures the core intent and most key parameters from the dependency_analysis; minor details (e.g., exact threshold numbers) may be softened but the overall workflow is fully recoverable
6-7: The fuzzy_description covers the main goal but misses or contradicts some important constraints, tools, or filtering steps described in the dependency_analysis
4-5: Significant misalignment, the fuzzy_description introduces unrelated requirements, drops critical steps, or implies a different workflow than what the dependency_analysis specifies
1-3: The fuzzy_description is essentially unrelated to the dependency_analysis, or actively contradicts the serial chain logic

Consider:

Does the fuzzy_description reflect the same user need and product domain as the dependency_analysis?
Are the key filtering criteria (category, brand, price range, rating thresholds, sentiment thresholds, keywords) from the dependency_analysis naturally embedded in the fuzzy_description?
Would an agent reading only the fuzzy_description be able to infer the correct serial tool chain described in the dependency_analysis?
Are there any extra requirements in the fuzzy_description that are absent from the dependency_analysis, or vice versa?


Output Format:
Provide scores and brief feedback in JSON format:

{{
  "solvability_score": <number 1-10>,
  "utility_score": <number 1-10>,
  "consistency_score": <number 1-10>,
  "solvability_feedback": "Brief explanation of solvability assessment",
  "utility_feedback": "Brief explanation of utility assessment",
  "consistency_feedback": "Brief explanation of consistency between dependency_analysis and fuzzy_description"
}}

\end{lstlisting}
\end{tcolorbox}

\twocolumn

\subsection{Error Analysis}

We conduct error analysis on GPT-5.4 trajectories in the L4 setting. 


\noindent\textbf{Dependency Reasoning Failure (54.1\%).}
This is the most common failure mode. 
It occurs when the agent fails to infer or respect dependencies among tools, such as executing downstream tools before prerequisite information is available, skipping required intermediate checks, or breaking multi-step dependency chains. 
These errors indicate that L4 tasks require more than isolated tool invocation; agents must maintain a coherent execution graph across multiple rounds.

\noindent\textbf{Target Retrieval Miss (53.6\%).}
In these cases, the agent completes a plausible tool-use trajectory but fails to retrieve or identify the target item. 
This often happens when intermediate evidence is incomplete, overly broad, or not properly propagated to later search and ranking steps. 
The failure suggests that even when tools are called successfully, the agent may still struggle to translate observations into the correct recommendation target.

\noindent\textbf{No Clear Failure / Correct Trajectory (18.5\%).}
A subset of trajectories does not exhibit an obvious planning or execution error under manual inspection. 
These cases are either correct or fail due to factors outside the annotated failure taxonomy, such as ambiguous user intent, incomplete tool returns, or borderline candidate relevance. 
We keep this category to avoid over-attributing errors to planning failures.

\noindent\textbf{Premature Termination (7.7\%).}
Here the agent stops before collecting sufficient evidence or before completing the required tool sequence. 
Typical cases include producing a final recommendation after only partial retrieval, omitting a necessary verification step, or stopping once a plausible candidate is found. 
This behavior reflects difficulty in judging when enough evidence has been accumulated.

\noindent\textbf{Unproductive Exploration (0.9\%).}
This category captures trajectories where the agent spends additional steps exploring irrelevant tools or candidates without improving the final decision. 
Although rare, such failures increase latency and may distract the agent from the original goal.

\subsection{Robustness of LLM-as-a-judge evaluation}
We assess how well LLM-based judgments align with human preferences using annotations from three independent reviewers. 
Each reviewer scores the output of each judge pipeline on a 3-point scale, where 0, 1, and 2 denote disagreement, partial agreement, and full agreement, respectively. 
We report the average score across reviewers and tasks. 
With prompt shuffling and score averaging, the human agreement score increases from 1.37 to 1.52 out of 2, suggesting improved alignment with human-perceived evaluation quality.

\end{document}